\documentclass[twocolumn,tighten,twocolappendix]{aastex701}

\usepackage{natbib}
\usepackage{graphicx} 
\usepackage{xcolor}
\usepackage{textgreek}
\usepackage[scr=boondoxo,scrscaled=1.05]{mathalfa}
\usepackage[utf8]{inputenc}
\usepackage[english]{babel}
\usepackage[caption=false]{subfig}
\UseRawInputEncoding
\usepackage{hyperref}
\hypersetup{
    unicode, 
    colorlinks=true,
    linkcolor=linkcolor,
    citecolor=linkcolor,
    filecolor=linkcolor,
    urlcolor=linkcolor,
}
\usepackage{color,colortbl}
\definecolor{linkcolor}{rgb}{0.0,0.3,0.5}
\usepackage{tensind}
\tensordelimiter{?}
\DeclareGraphicsExtensions{.bmp,.png,.jpg,.pdf}
\usepackage{verbatim}
\usepackage[normalem]{ulem}
\usepackage{orcidlink}
\usepackage{soul}
\usepackage{float}
\usepackage[flushleft]{threeparttable}
\usepackage{booktabs}
\usepackage{amsmath}

\newcommand{\M}{\mathcal{M}}
\graphicspath{ {./} }
\begin{document}

\title{Do we understand the star formation history of the universe?}
\author{Sahil Hegde\orcidlink{0000-0002-9370-8061}}
\altaffiliation{NSF Graduate Research \& NASA FINESST Fellow}
\affiliation{Department of Physics \& Astronomy, University of California, Los Angeles, 475 Portola Plaza, Los Angeles, CA 90095, USA}
\email[show]{sahil@astro.ucla.edu}
\author{Steven R. Furlanetto\orcidlink{0000-0002-0658-1243}}
\affiliation{Department of Physics \& Astronomy, University of California, Los Angeles, 475 Portola Plaza, Los Angeles, CA 90095, USA}
\email{sfurlane@astro.ucla.edu}
\author{Smadar Naoz\orcidlink{0000-0002-9802-9279}}
\affiliation{Department of Physics \& Astronomy, University of California, Los Angeles, 475 Portola Plaza, Los Angeles, CA 90095, USA}
\email{snaoz@astro.ucla.edu}

\begin{abstract}
     The evolving relationship between a galaxy's mass and star formation rate---the so-called `star-forming main sequence' (MS)---provides a critical benchmark for understanding star formation across time. Despite its fundamental importance, the observed main sequence remains subject to substantial systematic uncertainties in normalization, shape, and redshift evolution, and a longstanding discrepancy persists between the main sequence and its integral, the stellar mass function. We revisit the star-forming MS in the era of the James Webb Space Telescope by asking what star formation rates are \textit{required} by the stellar mass function to create a \textit{self-consistent} picture of galaxies across time. We fit compiled ground- and space-based measurements of star-forming and quiescent mass functions from $z=0.1-9$. By tracing galaxy growth histories through these mass functions, we present a statistically-robust \textit{inference} of the main sequence over $10^8 M_\odot \leq m_\star \leq 10^{11} M_\odot$, from the local universe to the first 500 Myr of cosmic history. Our procedure implies a main sequence that agrees with independent spectroscopic measurements of star formation rates from $z\sim 2-7$, is consistent with SED fitting-based analyses of photometric samples at $z\lesssim 3$, and aligns with theoretical models of galaxy evolution. However, we find that our MS differs from commonly-used `concordance' relations and thus caution against applications of these compilations without appropriately characterizing the underlying uncertainties. Finally, we explore the implications of our inferred main sequence for the galaxy-halo connection and star formation rate density, highlighting the need for further theoretical work to comprehensively understand the star formation history of the universe.
\end{abstract}

\section{Introduction}\label{sec:introduction}
Understanding the formation and evolution of galaxies across time has long been a pillar of modern astronomy. Enabled first by deep observations with the Keck and Hubble Telescopes in the late 20th century and expanded into the statistical arena with the first wide field galaxy surveys in the early 21st, observational studies of galaxy populations have sought to grow larger samples at progressively earlier epochs, in the hopes of piecing together a cohesive picture of the star formation history of the universe \citep{steidel_spectroscopy_1996, steidel_spectroscopic_1996, williams_hubble_1996, york_SDSS_2000, giavalisco_great_2004, scoville_cosmos_2007, grogin_candels_2011, madau_cosmic_2014}.\footnote{See also \cite{butcher_evolution_1978} for earlier studies of galaxy evolution more locally and \cite{sandage_optical_1978, gregory_coma_1978, huchra_survey_1983} for some of the first redshift surveys (and \citealp{giovanelli_redshift_1991, strauss_density_1995} for reviews of this pioneering work).} Population level inference began with these early statistical studies, and in turn, strong scaling relations between measured physical properties began to emerge (e.g., \citealp{cole_2dF_2001, tremonti_origin_2004, brinchmann_physical_2004}). 

\begin{figure*}
    \centering
    \includegraphics[width=\linewidth]{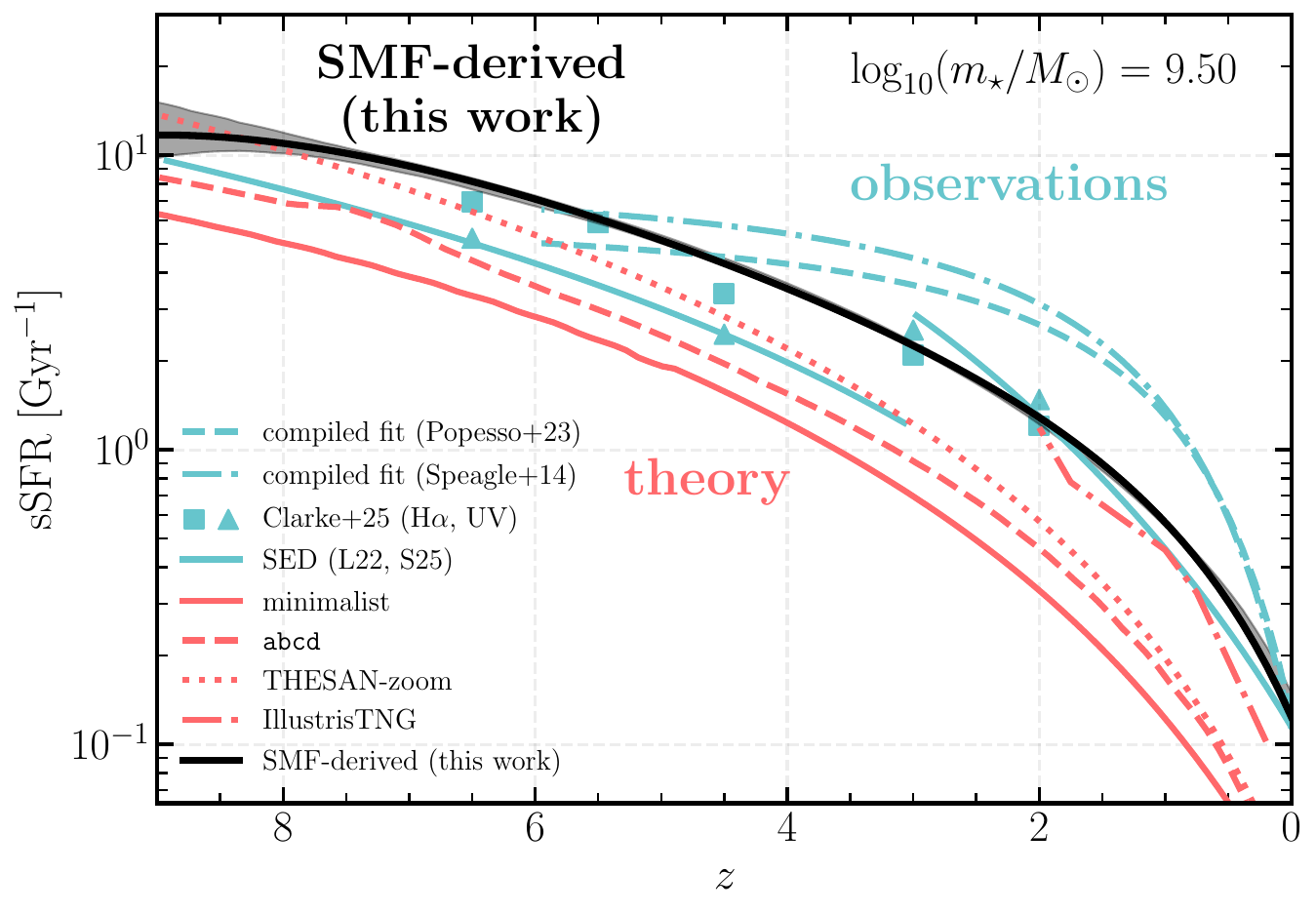}
    \caption{\textbf{JWST results straddle the divide between the star-forming MS inferred from a range of theoretical and observational studies, and the SFRs derived from the SMF are in strong agreement with these measurements.} The specific SFR evolution of $m_\star\sim 10^{9.5}M_\odot$ galaxies from $z\sim 0-9$. Theoretical star-forming MS estimates are shown in red, with different linestyles identifying different models: the minimalist model (\citealp{furlanetto_minimalist_2017}; solid), \texttt{abcd} model (\citealp{hegde_efficient_2025}, dashed), THESAN-zoom simulation (\citealp{mcclymont_thesan_2025d}, dotted), and IllustrisTNG simulation (\citealp{donnari_star_2019}, dot-dashed). Observational star-forming MS estimates are shown in blue, with different linestyles describing the different measurements: the latest compilation (\citealp{popesso_main_2023}; dashed), the preceding compilation (\citealp{speagle_highly_2014}; dot-dashed), JWST/NIRSpec-based measurements (\citealp{clarke_star_2025}; squares and triangles corresponding to H$\alpha$- and UV continuum-based SFR indicators, respectively), and the star-forming MS inferred using \texttt{Prospector} with pre- and post-JWST samples (\citealp{leja_new_2022, simmonds_bursting_2025}; solid). The star-forming MS inferred from the redshift evolution of the SMF is shown in black (see Sections~\ref{sec:continuity} and \ref{sec:SMF_MS}), with a shaded gray region corresponding to the 16th-84th percentile uncertainty interval derived from the joint posterior of the SMF fit parameters (discussed in Section~\ref{sec:obs_SMF} and Appendix~\ref{app:SMF_fit}).}
    \label{fig:sSFR}
\end{figure*}

One such relation --- and perhaps one of the most fundamental --- is the observed positive correlation between a galaxy's stellar mass and its derivative (the star formation rate, or SFR), and is accordingly denoted ``the main sequence of star formation'', or star-forming main sequence (star-forming MS;\footnote{Despite the unfortunate overlap in notation with the \textit{stellar} main sequence, we refer to this relation as the star-forming main sequence (or occasionally main sequence) to maintain consistency with past work.} e.g., \citealp{brinchmann_physical_2004, noeske_star_2007, whitaker_star_2012, whitaker_constraining_2014, whitaker_galaxy_2015, lee_turnover_2015, ilbert_evolution_2015, tomczak_SFR_2016, pearson_main_2018, davidzon_alternate_2018, iyer_sfr_2018, sandles_bayesian_2022, leja_new_2022, koprowski_charting_2024, simmonds_bursting_2025, clarke_star_2024, clarke_star_2025, cole_ceers_2025, perry_prevalence_2025, di_cesare_slope_2026, euclid_SFMS_2026}, and \citealp{speagle_highly_2014, popesso_main_2023} for recent compilations of the existing literature). Understood theoretically as a byproduct of hierarchical structure formation, which imprints a positive scaling between a galaxy's stellar mass, its host dark matter (DM) halo mass, and halo accretion rates (which in turn set galaxy baryon accretion rates), the normalization, shape, and evolution of the star-forming MS (and scatter around the mean relation) are thus sensitive probes of the physics governing gas flows, star formation, and feedback in our universe \citep{tacchella_confinement_2016, caplar_stochastic_2019, iyer_diversity_2020, tacchella_stochastic_2020, scholz_dark_2023, mcclymont_thesan_2025d, gui_episodic_2025, chu_galaxy_2026}. The observational robustness of this scaling implies that the vast majority of galaxies form much of their mass \textit{moving along} the star-forming main sequence \citep{leitner_last_2012} or \textit{passing through} it \citep{gladders_IMACS_2013, abramson_matching_2015, abramson_return_2016, corcho-caballero_single_2020, jain_self_2024}, and juxtaposing the movement of galaxies in this plane against theoretical models can shed light into the fundamental nature of galaxy evolution (for some recent work see e.g., \citealp{bouche_impact_2010, dekel_toy_2013, tacchella_confinement_2016, mcclymont_thesan_2025d, fortune_die_2025, kimmig_built_2026}). However, many studies have revealed a \textit{consistent} discrepancy between observed relations and theoretical models attempting to benchmark their physics against the measured trends, which should be self-consistent with halo growth histories and other global scaling relations (see e.g., Figure~\ref{fig:sSFR} and \citealp{dave_galaxy_2008, dave_mufasa_2016, mitchell_evolution_2014, sparre_star_2015, somerville_physical_2015, donnari_star_2019}).

Despite the wealth of observational samples and attention that measuring the star-forming main sequence has garnered, there is also a longstanding systematic uncertainty in the observed shape and normalization of the relation (Figure~\ref{fig:sSFR} and e.g., \citealp{speagle_highly_2014, katsianis_high_2020, leja_new_2022, popesso_main_2023}). In particular, it has been shown that the former of these phenomena can be understood as a reflection of differences in selection criteria (i.e., how star-forming galaxies are identified in a survey) and in how physical properties are inferred (through calibration differences in SFR or stellar mass indicators, which can be especially uncertain for the former;\footnote{For example, \cite{pacifici_art_2023} find that while stellar masses are fairly robust across a wide range of SED fitting assumptions, SFRs can be significantly more uncertain, because, in addition to their dependence on the adopted choice of star formation history (and the priors applied therein; SFH), the inferred SFRs are also highly sensitive to the assumed dust law.} see \citealp{conroy_propagation_2009, conroy_propagation_2010, conroy_propagation_2010b, mitchell_how_2013, mobasher_critical_2015, pacifici_art_2023}). \cite{speagle_highly_2014} (and later \citealp{popesso_main_2023}) carried out a comprehensive accounting of these effects and demonstrated that under a mutual recalibration of the measured data, the systematic variance can be tightened and a consistent star-forming main sequence emerges. While this is encouraging, it does not ensure that the resulting star-forming MS is the \textit{true}, fundamental relation, and is not simply a proxy for biases in our inference of physical parameters. 

In this spirit, \cite{leja_new_2022} revisited the observed star-forming main sequence with a novel approach to inferring galaxy properties using the Bayesian spectral energy distribution (SED) modeling framework \texttt{Prospector} \citep{leja_deriving_2017, johnson_stellar_2021}. Leveraging the ability of this framework to simultaneously vary (and constrain) star formation histories (SFHs), metallicities, and dust attenuation, \cite{leja_older_2019, leja_new_2020, leja_new_2022} suggested that preceding calibrations underestimated stellar masses (by 0.2-0.3 dex) and overestimated SFRs. \cite{leja_new_2022} demonstrated that a star-forming main sequence derived from galaxy samples at $z\sim 0.2-3$ with these uncertainties in mind could additionally reconcile discrepancies with theoretical models (though see also \citealp{baldwin_comparison_2018, han_comprehensive_2019, leja_older_2019, van_mierlo_no_2023, pacifici_art_2023} for a comparison of different stellar population synthesis models and SED fitting codes, and a discussion of the biases that can arise).

In tandem, the launch of the James Webb Space Telescope (JWST) has reinvigorated studies of galaxies to exceptionally early times and enabled a characterization of galaxy populations with exquisite sensitivity into the infrared (IR). While the discoveries at the earliest, previously unexplored epochs ($z\gtrsim 10$) have garnered significant attention, at comparatively more modest redshifts, JWST is filling in our understanding of the HST-dark, red galaxy population, for example uncovering and quantifying the contribution of dusty star-forming galaxies to the global star formation rate density (SFRD; see e.g., \citealp{gottumukkala_unveiling_2024, williams_galaxies_2024, de_graff_rubies_2025, barrufet_quiescent_2025, barrufet_strength_2026, cooper_rubies_2025, baker_exploring_2025, gentile_going_2025, sun_spectroscopic_2025}). In addition to uncovering a vast population of previously unexplored sources, the higher sensitivity and redder spectral coverage of JWST enable more robust physical characterization of galaxies by probing their rest-IR and rest-optical SEDs across time (see \citealp{shapley_JWST_2023, shapley_aurora_2025, sanders_excitation_2023, sanders_direct_2024, sanders_aurora_2026, pahl_spectroscopic_2025, clarke_star_2025, acharya_comparing_2026, karthikeyan_balmer_2026} for some examples).

Noting the systematic uncertainties in the `measured' star-forming MS and the power of JWST to provide a more \textit{complete} and \textit{physically robust} census of star formation across time, in this work we revisit the star-forming main sequence with a different approach: inference via the redshift evolution of the galaxy stellar mass function (SMF). Because galaxies grow due to star formation (among other things), the instantaneous state of the SMF represents the galaxy population at a particular epoch and the time evolution is essentially an integral of the star-forming main sequence. In this sense, if our understanding of galaxies is complete and our measured scaling relations `correct,' then the star-forming main sequence and mass function should be consistent with one another. Put another way, \textit{the evolution of the SMF provides a phenomenological `test' of the self-consistency of our observed star-forming main sequence}. 

This approach has been employed in the past to evaluate galaxy star formation rates in the pre-JWST era \citep{bell_star_2007, weinmann_fundamental_2012, ilbert_mass_2013, leja_reconciling_2015, davidzon_alternate_2018} and has been implicitly applied to quantify the effects of other processes on galaxy evolution as well (such as mergers and quenching; e.g., \citealp{drory_contribution_2008, peng_mass_2010a, peng_mass_2012, peng_mass2014a, behroozi_average_2013, behroozi_universemachine_2019, wang_relating_2023}). For example, \cite{leja_reconciling_2015} leveraged this procedure to demonstrate that an extrapolation of the then-latest star-forming MS measurements resulted in a significant overprediction of the resulting SMF when integrated over time, concluding that the low-mass star-forming MS must deviate from a naive power law extrapolation of the high-mass relation (a feature which was then borne out in subsequent studies of the star-forming MS; \citealp{whitaker_constraining_2014, lee_turnover_2015, tomczak_SFR_2016, leja_new_2022, popesso_main_2023}). In fact, we find that a similar analysis applied to the latest SMF measurements with the most recent `concordance' star-forming MS compilations yields an analogous overprediction, suggesting that biases in this fundamental relation may persist (see Appendix~\ref{app:evolving_SMF} for a demonstration of this result).

Therefore, in this work, we use revised estimates of the SMF over $0.1 < z < 9$ to invert this phenomenological test and \textit{infer} mean galaxy star formation rates across time from derivatives of the mass function. Because a careful accounting of the evolution of both the star-forming and quiescent galaxy populations across time is necessary to accurately infer the true growth of the mass function due to star formation, we emphasize that such analysis is only newly possible in light of the samples produced by JWST. We combine SMF measurements from the local universe \citep{moustakas_primus_2013, xu_pac_2025} with recent pre- and post-JWST surveys \citep{stefanon_galaxy_2021, weaver_cosmos2020_2023, shuntov_cosmos-web_2025, shuntov_stellar_2026} to survey the evolving galaxy population across time, infer the star-formation histories required by this stellar mass buildup, and compare the resulting star-forming main sequence to the existing literature.

Our paper is organized as follows. In Section~\ref{sec:star-forming MS_overview} we introduce a collection of representative star-forming main sequences from the literature to establish a baseline for our comparison. In Section~\ref{sec:SMF_evolution}, we detail our approach for deriving SFRs from the evolution of the SMF and discuss our characterization of confounding effects such as mergers and quenching (a summary of which is provided in Section~\ref{sec:procedure}). In Section~\ref{sec:obs_inference}, we construct a continuous analytical fit to a compilation of observed SMF measurements over time and in Section~\ref{sec:SMF_MS} we present the star-forming MS that results from applying our procedure to this data. Next, in Section~\ref{sec:discussion} we juxtapose this star-forming main sequence against pre-existing samples and discuss the implications of our findings. Finally, in Section~\ref{sec:conclusion}, we conclude. We note that the results of our fitting procedures are summarized in Appendices~\ref{app:SMF_fit} and \ref{app:star-forming MS_fit}, and we provide a publicly available Python implementation of these fits to enable a direct comparison with our results.\footnote{\href{https://github.com/hegdesahil/SMF_SFMS}{https://github.com/hegdesahil/SMF\_SFMS}}

In any discussion of the DM halo population, we use the Sheth-Tormen halo mass function \citep{sheth_ellipsoidal_2001} and halo accretion rates from the Millennium simulation \citep{fakhouri_merger2010}. Whenever necessary we use a standard $\Lambda$CDM cosmology consistent with the latest results from \cite{Planck21}. All stellar masses used in this work have been converted to a \cite{Chabrier03} initial mass function (IMF) for consistency.

\section{The star-forming star-forming main sequence} \label{sec:star-forming MS_overview}
Before discussing our new star-forming main sequence inference, we establish an observational benchmark and reiterate the existing tensions in this quantity with a range of representative estimates from the literature, both theoretical and observed. We show the specific SFR (${\rm sSFR}\equiv {\rm SFR}/m_\star$) over time implied by these relations in Figure~\ref{fig:sSFR} and the full star-forming main sequences in Figure~\ref{fig:MS_comparisons} (though both of these Figures also show the SMF-derived relation from this work, we defer a detailed description of our methodology to Section~\ref{sec:SMF_evolution}).

\subsection{Observational results}\label{sec:observed_SFMS}
We begin by summarizing a few recent \textit{observed} star-forming MS measurements.

\textbf{Observed compilations:} 
The most recent fits to compiled star-forming MS measurements are described in \cite{speagle_highly_2014, popesso_main_2023}. 

Relying on a collection of star-forming main sequence measurements derived from a wide range of SFR indicators and assumptions about metallicity, dust, IMF, etc., \cite{speagle_highly_2014} infer a `concordance' star-forming main sequence evolving out to $z\lesssim 6$. They account for discrepancies in the calibrations, conventions, and other systematic uncertainties, bringing the measurements into a consistent basis, and fit the compiled measurements to the following functional form:
\begin{equation}
    \label{eq:speagle_fit}
    \log_{10}\dot{m}_\star(\M, t) = a(\M)t + b(\M),
\end{equation}
where $t$ is the Hubble time in Gyr and $a(\M)$ and $b(\M)$ are linear functions of the logarithmic mass. This MS is shown with a blue dot-dashed line in Figure~\ref{fig:sSFR}. 

\cite{popesso_main_2023} carry out a similar exercise, updating the \cite{speagle_highly_2014} estimate with MS measurements from the intervening decade. By including these measurements along with the original \cite{speagle_highly_2014} result, they find that the MS can be well-described with a broken power-law:
\begin{equation}\label{eq:p23_fit}
    \log_{10}\dot{m}_\star = a_0 + a_1t -\log_{10}\bigg(1 + \bigg(\frac{m_\star}{10^{a_2+a_3 t}}\bigg)^{-a_4}\bigg)\ ,
\end{equation}
This results in a star-forming MS with a low-mass logarithmic slope of 1 and a high-mass flattening dictated by two key parameters: a peak SFR and turnover mass, both of which evolve log-linearly with cosmic time \citep{lee_turnover_2015, tomczak_SFR_2016}. This star-forming main sequence is shown with a blue dashed line in Figures~\ref{fig:sSFR} and \ref{fig:MS_comparisons}. Despite the promise of these reconciliations, these results do not guarantee that the resulting star-forming main sequence reflects the ground truth---as is often assumed in the literature---and indeed can be subject to biases in the chosen basis itself \citep{katsianis_high_2020, popesso_main_2023}. 

\cite{popesso_main_2023} note that this inferred star-forming MS reproduces a longstanding discrepancy with the theoretical literature (as we demonstrate in Figure~\ref{fig:sSFR}). Namely, at $z\lesssim 3-4$, theoretical models --- which almost universally derive SFRs in a manner tied to the redshift evolution of DM halo accretion rates --- predict a steeper redshift evolution of the star-forming MS than the observed results suggest (until $z\sim 0$, when the results converge; see e.g., their Figure 7).

\textbf{SED fitting-based star-forming MS:} Though the concordance star-forming main sequence presented in \cite{popesso_main_2023} fits the available data well, contemporaneous refinements in SED fitting techniques and the advent of JWST suggest that modifications to the base model may be warranted. Based on the non-parametric SFHs fit using the SED fitting routine \texttt{Prospector}, \cite{leja_new_2020, leja_new_2022} argue that previous stellar mass estimates at $z\lesssim 3$ derived from UV+IR photometry have been systematically underestimated by $0.2-0.3$ dex at all masses, and that SFRs may, in some cases, be overestimated (see also \citealp{leja_older_2019} for a detailed discussion). Correcting these biases shifts galaxies to higher stellar masses and lower SFRs in the star-forming MS parameter space (i.e., to the `right' and `downward'). \cite{leja_new_2022} revise the inference of the star-forming MS in light of these corrections and argue that the resulting relation resolves discrepancies with theoretical models.

Extending this methodology to higher redshift, JWST photometry analyzed with \texttt{Prospector} at $z\sim 3-9$ suggests a lower normalization and somewhat shallower late-time evolution than suggested by previous observations of the star-forming MS \citep{simmonds_bursting_2025}. Because of the consistency in analysis framework and minimal overlap in redshift coverage, we combine these results as a single solid blue curve in Figures~\ref{fig:sSFR} and \ref{fig:MS_comparisons}.

\textbf{H$\alpha$-based star-forming MS:} The redshift evolution of the star-forming MS has also been explored with JWST spectroscopy \citep{clarke_star_2025, lam_JADES_2026}. The redder spectral coverage of JWST's NIRCam and NIRSpec instruments improves stellar mass estimates, by providing more complete inputs for SED fitting, and star formation rates, through direct measurements of H$\alpha$ luminosities to $z\lesssim 7$ with updated metallicity calibrations (e.g., \citealp{shapley_JWST_2023}). Leveraging this power, \cite{clarke_star_2025} characterize the star-forming MS at discrete redshift intervals between $z\sim 1.4-7$ and report the relation inferred based on a galaxy's H$\alpha$ flux and $1600\AA$ UV luminosity. We display these results with blue squares and triangles in Figure~\ref{fig:sSFR}, respectively.

\subsection{Theoretical models}\label{sec:star-forming MS_theory}
The main sequence is also a natural prediction of most theoretical models. Here we summarize a few representative estimates in order of increasing model complexity.

\textbf{The minimalist MS:} The simplest theoretical models for galaxy formation associate steady-state star formation rates with DM halo accretion rates, with the scaling and normalization regulated by stellar feedback \citep[e.g., ][]{bouche_impact_2010, dave_analytic_2012, dekel_toy_2013, lilly_gas_2013, tacchella_confinement_2016, sun_constraints_2016, furlanetto_minimalist_2017, mirocha_effects_2020, furlanetto_bursty_2022}. In this limit, the redshift evolution of the star-forming MS is broadly set by the redshift evolution of mean DM halo accretion rates, and the slope is set by the details of stellar feedback (e.g., energy- vs. momentum-regulated SN feedback introduce different halo mass dependencies for the feedback strength). 

One such model is the minimalist model \citep{furlanetto_minimalist_2017}. In this framework, SFRs derived from DM halo accretion rates under energy- and momentum-regulated feedback reproduce observed galaxy luminosity functions at $z\sim 6-10$. Here, we extend this model to $z\sim 0$, holding the feedback prescription fixed. We show the energy-regulated result in Figures~\ref{fig:sSFR} and \ref{fig:MS_comparisons} as a solid red line.

\textbf{The \texttt{abcd} star-forming MS:} Building from the minimalist model, which predicts the \textit{equilibrium} SFR of a host DM halo, we also consider the \texttt{abcd} galaxy formation model, which traces cycles of `bursty' star formation driven by delayed stellar feedback \citep{furlanetto_bursty_2022, hegde_self-consistent_2023, hegde_efficient_2025}. We again fix the feedback model to a $z\sim 6-10$ UVLF calibration and extend the star formation histories to $z\sim 0$ to quantify the effect of introducing a simple burst prescription on the inferred mean SFRs. We show the \texttt{abcd} star-forming MS with a dashed red line in Figures~\ref{fig:sSFR} and \ref{fig:MS_comparisons}.\footnote{To account for galaxy quenching, which is \textit{not} included in the base \texttt{abcd} model --- and itself is subject to significant uncertainty --- we artificially flatten the inferred star-forming MS at $m_\star \gtrsim 10^{11} M_\odot$ at all redshifts.}

\textbf{IllustrisTNG:} The star-forming MS can also be inferred from a cosmological simulation of galaxy formation such as IllustrisTNG, which captures stellar and AGN feedback alongside the complex gas flows and other dynamical effects that modulate galaxy evolution \citep{springel_first_2018}. \cite{donnari_star_2019} carry out a mock observational selection of star-forming galaxies in the IllustrisTNG 100 and 300 Mpc$^3$ boxes at $z\lesssim 2$ to estimate the simulated star-forming MS. While the simulation similarly reproduces a main sequence that falls in the ballpark of observations at $z\sim 0$, the inferred relation between SFR and stellar mass grows more slowly than most observed star-forming MS compilations, an example of the discrepancy discussed in \cite{popesso_main_2023} (see also Figure~\ref{fig:sSFR}). The IllustrisTNG star-forming MS is shown as a dot-dashed red line in Figures~\ref{fig:sSFR}.

\textbf{THESAN-zoom:} The THESAN simulation builds on the IllustrisTNG galaxy formation model with a radiative transfer model designed to study the process of reionization and its effect on galaxies at $z\gtrsim 5$ \citep{kannan_introducing_2022b}. In a high-resolution extension to this suite, the THESAN-zoom simulations explore the multiphase structure of the ISM to model the star formation process in more detail \citep{kannan_introducing_2025}. \cite{mcclymont_thesan_2025d} focus this analysis to study the evolution of galaxies onto and around the star-forming MS as they proceed through cycles of stochastic star formation. We display the star-forming MS inferred this way as a dotted red line in Figures~\ref{fig:sSFR} and \ref{fig:MS_comparisons}.

Stochastic star formation in simulations proceeds as a result of internal and external variability in gas inflows, driven by variable accretion rates, mergers, and stellar feedback \citep{caplar_stochastic_2019, tacchella_stochastic_2020, iyer_diversity_2020, furlanetto_bursty_2022, sun_seen_2023, pallottini_stochastic_2023, gelli_impact_2024, gelli_temporarily_2025, iyer_stochastic_2024, mcclymont_thesan_2025d, munoz_relatively_2026}. As a result, the simulated star-forming MS reflects the SFRs expected in a regime with a more `realistic' degree of burstiness (relative to the simpler feedback-driven bursts of the \texttt{abcd} model). Because of the limited zoom volume employed in this case, the THESAN-zoom star-forming MS is limited to masses below $m_\star\lesssim 10^{10} M_\odot$ and is derived from $3\leq z\leq 12$, so any predictions at lower redshifts are extrapolations of the inferred relations.

Juxtaposing these theoretical and observational star-forming main sequence estimates in Figure~\ref{fig:sSFR}, it is evident that (1) observational and modeling systematics can introduce variation in the time evolution and normalization of the measured star-forming main sequence, and (2) theoretical models consistently \textit{underpredict} most observed relations, albeit with a (roughly) mutually consistent redshift evolution (matching the $\dot{m}_h \propto (1+z)^{5/2}$ scaling of mean DM halo accretion rates). These discrepancies motivate an alternative approach that does not rely on direct measurements of galaxy star formation rates. Instead, we infer the star-forming main sequence from the observed evolution of the stellar mass function, using the fact that the buildup of stellar mass encodes the integrated history of galaxy growth.

\section{Inferring the star-forming MS from the evolution of the SMF}\label{sec:SMF_evolution}
Because the SMF records the cumulative buildup of stellar mass across cosmic time, its evolution provides an independent constraint on the average star formation histories of galaxies. Recovering the star-forming MS from the SMF therefore amounts to identifying the star formation rates required to transform the observed mass function from one epoch to the next.

However, stellar mass growth is not exclusively driven by star formation. The evolution of the SMF is governed by three primary processes: star formation, mergers, and quenching. Star formation shifts galaxies toward higher stellar masses, mergers redistribute galaxies across the mass function, and quenching halts further stellar growth. Although star formation dominates the evolution of the SMF over much of cosmic history, accurately inferring the star-forming MS requires accounting for the effects of all three processes. In the following, we present the mathematical framework underlying this inference and describe how each process is incorporated.

\subsection{A continuity equation for galaxies}\label{sec:continuity}
Assuming continuous star formation and isolated evolution (i.e., no mergers), a population of galaxies \textit{preserves their mass rank order} and thus the total number density of galaxies above an (evolving) stellar mass is conserved:
\begin{equation}
    \label{eq:total_deriv}
    \frac{dn(>m_\star(t), z)}{dt} = 0 \ .
\end{equation}
Here, $n$ represents the cumulative number density of galaxies and is defined with respect to the (logarithmic) stellar mass function $\Phi$ (i.e., $[\Phi] = M_\odot\ {\rm dex^{-1}}$) --- i.e., $n(>m_\star,z) = \int_{\M}^{\infty}\Phi d\M'$, where $\M\equiv \log_{10}m_\star$ is the logarithmic mass. Expanding the total derivative, this equation becomes
\begin{equation}
    \label{eq:cum_continuity_conserved}
    \frac{dn(>m_\star, z)}{dt} = \frac{\partial n}{\partial t} + \frac{1}{\ln 10}\frac{\Phi \langle\dot{m}_\star\rangle}{m_\star} = 0 \ ,
\end{equation}
where the second term results from applying the fundamental theorem of calculus to the cumulative mass function integral. Taking the derivative with respect to the logarithmic mass variable, we recover the usual continuity equation applied to the stellar mass function:
\begin{equation}
    \label{eq:continuity_conserved}
    \frac{\partial \Phi(m_\star, t)}{\partial t} + \frac{1}{\ln10}\frac{\partial}{\partial \M}\bigg[\frac{\Phi \langle\dot{m}_\star\rangle}{m_\star}\bigg] = 0 \ ,
\end{equation}
where $\langle\dot{m}_\star\rangle$ represents the average stellar mass `velocity' of galaxies in this sample (i.e., the average star formation rate of sources described by the mass function $\Phi$). In this limit, the absolute rate at which the abundance of galaxies of a particular mass changes is  equal to the average mass flow rate in and out of that bin due to star formation. Importantly, this formalism implies that the evolution of the stellar mass function does not depend on the scatter in star formation rates at a given stellar mass (i.e., the scatter in the star-forming MS), assuming that true mean of the distribution is known \citep{leja_reconciling_2015}. In this work we leverage the converse of this result: that is, the star formation rates inferred from the evolution of the stellar mass function represent the average star formation rate of the star-forming galaxy population, or the intrinsic star-forming main sequence. 

However, though it is conventional to define the SMF in terms of the \textit{surviving} stellar mass of a galaxy (i.e., including the effects of stellar mass loss), the star-forming MS is often defined in terms of the \textit{integrated} stellar mass of a galaxy.\footnote{We note that this convention is not necessarily adopted in all definitions of the star-forming MS. We therefore encourage care in keeping track of which mass is being reported when interpreting a measured star-forming MS in the context of the SMF or other galaxy properties, especially when comparing to theoretical models, which often focus on the integrated stellar mass. This issue becomes particularly difficult to quantify when leveraging compilations of the star-forming MS, such as \cite{speagle_highly_2014} or \cite{popesso_main_2023}, which may include star-forming MS measurements using both definitions.} These differ by the fraction of material returned to the ISM due to mass loss during stellar evolution, for which we assume a constant $R\approx 0.36$ (i.e., stellar mass loss is assumed to be effectively instantaneous, which is a reasonable assumption for timescales $>100$ Myr), consistent with previous work \citep{leitner_fuel_2011, leja_reconciling_2015, tacchella_confinement_2016}. Thus, the star-forming MS is defined as
\begin{equation}
    \label{eq:star-forming MS_from_Mstaravg}
    \dot{m}_\star^{\rm SFMS} \equiv \frac{\langle \dot{m}_\star\rangle}{1-R}\ .
\end{equation}

This isolated evolution limit (i.e., setting $dn/dt = 0$) has featured in many previous studies attempting to model galaxy growth histories by matching the stellar mass function across time \textit{at a constant comoving number density} \citep[e.g.,][]{brown_evolving_2007, brown_red_2008, van_dokkum_growth_2010, van_dokkum_assembly_2013, papovich_rising_2011, papovich_zfourge_2015, ownsworth_minor_2014, leja_reconciling_2015, davidzon_alternate_2018}. While it has been shown to provide a decent approximation to average simulated growth histories in some cases (e.g., \citealp{jaacks_connecting_2016, clauwens_large_2016, wang_relating_2023}), this procedure ignores known non-conservative processes such as mergers and quenching that can shuffle the mass rank order of galaxies and thus modulate the inferred SFRs as well.

\subsection{Relaxing the assumptions}\label{sec:nonconservative_effects}
Mergers and quenching do not conserve the number of galaxies in the stellar mass function --- the former lead to absorption of low mass galaxies into more massive systems, and the latter can remove galaxies from the star-forming population entirely. In this section, we will revisit the continuity equation introduced in Eq.~(\ref{eq:continuity_conserved}), including these processes, with a specific focus on the \textit{star-forming stellar mass function}.

With non-conservative effects included, Eq.~(\ref{eq:cum_continuity_conserved}) becomes
\begin{equation}
    \label{eq:cum_continuity}
    \frac{dn_{\rm SF}}{dt} = \frac{\partial n_{\rm SF}}{\partial t} - \frac{1}{\ln10}\bigg[\frac{\Phi_{\rm SF} \langle\dot{m}_\star^{\rm tot}\rangle}{m_\star}\bigg] = -\mathcal{Q} - \mathcal{A} \ ,
\end{equation}
where now we have introduced the subscript $_{\rm SF}$ to make explicit that we are considering the star forming galaxy population alone and the average stellar mass growth of galaxies depends both on star formation and mergers: $\langle\dot{m}_\star^{\rm tot}\rangle = \dot{m}_\star^{\rm SF} + \dot{m}_\star^{\rm merg}$.\footnote{Note that here the return fraction correction is applied only to the first term in the sum to infer the star-forming MS from the average SFR. In other words, $\dot{m}_\star^{\rm SFMS} = \dot{m}_\star^{\rm SF}/(1-R) =(\langle\dot{m}_\star^{\rm tot}\rangle - \dot{m}_\star^{\rm merg})/(1-R)$.} The two terms on the RHS quantify the effects of quenching ($\mathcal{Q}$, as such galaxies leave the star-forming population) and mergers (denoted by $\mathcal{A}$ for `absorption') on modifying the galaxy mass rank order. Because we are working in the space of star-forming galaxies alone, the star formation rates inferred from this procedure represent the average SFRs of star-forming galaxies (i.e., the star-forming MS; cf. Appendix~\ref{app:evolving_SMF} and \citealp{leja_reconciling_2015} for procedural differences in the context of the full galaxy population).

For completeness, the differential form of this equation is as follows
\begin{equation}\label{eq:continuity}
    \frac{\partial \Phi_{\rm SF}(m_\star, t)}{\partial t} + \frac{1}{\ln10}\frac{\partial}{\partial \M}\bigg[\frac{\Phi_{\rm SF} \langle\dot{m}_\star^{\rm tot}\rangle}{m_\star}\bigg] = -\mathscr{q} - \mathscr{a}\ ,
\end{equation}
though in what follows we will work in terms of cumulative number densities because it is more straightforward to characterize mergers in that case.

\textbf{Quenching:} The first of the `sink' terms in the continuity equation is the quenching term $\mathcal{Q}$. Because we are working in the space of star forming galaxies, the process of galaxy quenching removes galaxies from the population. We quantify the rate at which galaxies are removed from the star forming sample by evaluating the time derivative of the cumulative quiescent stellar mass function $n_{\rm Q}$:
\begin{equation}
    \label{eq:quench_rate}
    \mathcal{Q} \equiv \frac{\partial n_{\rm Q}(>m_\star, z)}{\partial t}\ .
\end{equation}

Mergers similarly scramble the mass rank order of galaxies through absorption of low mass systems by more massive galaxies. This shifts massive galaxies into higher mass bins through the addition of less massive systems and simultaneously removes galaxies, thus changing the number densities. 

\textbf{Merger-driven mass growth:} The former of these effects is the mass growth due to minor mergers, which introduces an additional contribution to $\dot{m}_\star^{\rm tot}$ in Eq.~(\ref{eq:cum_continuity}) associated with $\dot{m}_\star^{\rm merg}$. To estimate this effect, we begin with the fitting function for the number of DM halo mergers as a function of halo mass, mass ratio $\xi = m/M$, and redshift, $dN/d\xi dz(M,\xi, z)$, based on the Millennium-I and II simulations \citep{fakhouri_merger2010}. With this parameterization, the mass growth rate due to mergers is
\begin{equation}\label{eq:mdot_merge}
    \begin{split}
        \dot{m}_{\star}^{\rm merg}(m_\star,z) &= \frac{dz}{dt}\int_{\xi_{\rm min}}^1 \xi m_\star/f_\star \\
    &\qquad\qquad\times\frac{dN}{dzd\xi}\Big(m_\star/f_\star, \xi, z\Big)d\xi\ ,
    \end{split}
\end{equation}
where we have introduced the stellar-to-halo mass relation (SHMR) $m_\star \equiv f_\star m_h$ to make manifest that we are estimating the \textit{stellar} mass growth rate due to mergers. We derive an approximate form of this relation by abundance matching of the stellar and halo mass functions:
\begin{equation}
    \label{eq:SHMR}
    \int_{\log_{10}m_h}^\infty n(\M_h')d\M_h' = \int_{\log_{10}m_\star}^\infty \Phi_{\rm tot}(\M_\star')d\M_\star'\ ,
\end{equation}
where $\log_{10}m_\star = \log_{10}f_\star m_h$. We choose $\xi_{\rm min} = 0.1$ here but note that the growth rate due to mergers minimally depends on this choice. This estimate results in merger-driven mass growth rates that increase with galaxy stellar mass, consistent with the results of the \cite{guo_galaxy_2013} semi-analytic model shown in \cite{leja_reconciling_2015}. In particular, merger-driven mass growth is subdominant to star formation in all galaxies below $m_\star \lesssim 10^{11} M_\odot$, with a slight dependence on redshift. 

\textbf{Merger-driven absorption:} The latter effect of mergers is to remove low mass systems. Because low-mass systems get absorbed by more massive galaxies, the progenitor tracks of individual galaxies deviate from the constant number density assumption described in Sec.~\ref{sec:continuity}. Rather than evolving at a fixed abundance as might be expected for smooth, star formation-driven growth, the most massive progenitor of a given galaxy tends to be found in a \textit{higher abundance bin} than its descendant, albeit with significant scatter in this relation (due to scatter in galaxy merger histories). 

\cite{behroozi_using_2013} quantify this effect using simulated DM halo growth histories in a procedure we briefly summarize here for completeness. First, \cite{behroozi_using_2013} associate galaxy masses with halo masses by abundance matching an observed $z\sim 0$ galaxy stellar mass function with halo mass functions derived from the Bolshoi simulation. Next, by tracing the range of cumulative number densities of the progenitors associated with a sample of $z\sim 0$ galaxy masses, they recognize that the median cumulative number density of galaxy progenitors evolves log-linearly with redshift:
\begin{equation}
    \label{eq:dndz_merger}
    \frac{d\log_{10}n(>m_\star,z)}{dz} \approx \alpha(m_\star)\ ,
\end{equation}
where $\alpha$ is a constant that depends mildly on galaxy mass and the details of the simulated halo growth histories \citep{behroozi_using_2013}.

This parameterization improves on the aforementioned constant cumulative number density approximation considerably and was shown to be remarkably effective in reproducing the median growth histories of simulated galaxies \citep[e.g.,][]{torrey_analysis_2015, jaacks_connecting_2016, clauwens_large_2016, torrey_forward_2017, wellons_improved_2017, davidzon_alternate_2018, wang_relating_2023}. However, while \cite{behroozi_using_2013} derived this relation for the \textit{total} galaxy stellar mass function, it was subsequently found that the stellar mass evolution (and thus cumulative number density evolution) depends significantly on the star formation rate of the $z\sim 0$ descendant galaxy \citep{clauwens_large_2016, wang_relating_2023}. In particular, $z\sim 0$ quiescent galaxies tend to have more massive progenitors ($\alpha\sim 0$ in Eq.~(\ref{eq:dndz_merger})) than their star-forming counterparts ($\alpha\sim 0.16$ for $m_\star \lesssim 10^{11.2} M_\odot$ and $\sim 0.22$ for $m_\star > 10^{11.2} M_\odot$).

We apply this parameterization to quantify the cumulative number density change due to mergers. Recasting Eq.~(\ref{eq:dndz_merger}) in terms of cosmic time, we have
\begin{equation}
    \label{eq:merger_absorption}
    \mathcal{A} \equiv \frac{\partial n_{\rm SF}}{\partial t}\Bigg|_{\rm merg} = n_{\rm SF} \ln 10 \frac{dz}{dt}\alpha(m_\star)\ .
\end{equation}
Note that with $\alpha$ defined as we have, this parameterization is only appropriate for tracing the progenitors of $z\sim 0$ galaxies because forward and backward number density evolution is asymmetric \citep{behroozi_using_2013, torrey_forward_2017}.\footnote{Because low mass galaxies are more abundant than more massive systems, faster growth trajectories are overrepresented in the descendant median number density evolution. Reconciling the difference between the two directions requires a convolution of the probability that a galaxy with an initial number density $n_0$ will evolve into a galaxy with a number density $n_f$ with the relative abundances of each of those $n_0$ galaxies. In practice, beginning at $z_i>0$ and integrating galaxy tracks forward in time (i.e., looking at galaxy \textit{descendants}) amounts to requiring a different calibration that is not well quantified across our redshift range of interest, so we limit our analysis to the backward tracking direction \citep{torrey_forward_2017}.}

\subsection{The star-forming MS inference procedure}\label{sec:procedure}
Altogether, our Eulerian evolution equation for the star-forming stellar mass function can be summarized as 
\begin{equation}
    \label{eq:cum_sf_continuity_full}
    \frac{\partial n_{\rm SF}}{\partial t} =  \frac{1}{\ln10}\bigg[\frac{\Phi_{\rm SF} \langle\dot{m}_\star^{\rm tot}\rangle}{m_\star}\bigg] - \frac{\partial n_{\rm Q}}{\partial t} - n_{\rm SF} \ln 10 \frac{dz}{dt}\alpha\ .
\end{equation}
Because this equation involves both the stellar mass function $\Phi$ and its integral $n(>m_\star)$, and our merger-driven absorption criterion Eq.~(\ref{eq:merger_absorption}) is only well-defined for tracing galaxy progenitors from $z\sim 0$, in practice we infer the star-forming MS by tracing the growth histories of individual galaxies satisfying this evolution. This essentially amounts to solving the PDE by tracing the mass evolution along characteristic curves anchored to the galaxy mass at $z\sim 0$.

Our procedure is summarized as follows:
\begin{enumerate}
    \item Given star-forming and quiescent stellar mass functions $\Phi_{\rm SF}(m_\star, z)$ and $\Phi_{\rm Q}(m_\star,z)$, construct a grid of final galaxy masses $\{m_{\star,0}^{i}\}$ at $z=0$ and initialize their number densities as $\{n_{\rm SF}(>m_{\star,0}^{i}, 0)\}$.
    \item Step backward to an earlier time $z_j>0$ and compute the median progenitor abundance using Eqs.~(\ref{eq:quench_rate})-(\ref{eq:dndz_merger}); i.e., 
    \begin{align}\label{eq:new_abund}
    \begin{split}
    n&_{\rm SF}(>m_{\star,0}^i,0)\mapsto \\ &\big[(n_{\rm SF}(>m_{\star,0}^i,0) + \Delta n_{\rm Q}(>m_{\star,0}^i, 0)\big]\times 10^{\alpha \Delta z}\ ,
    \end{split}
    \end{align}
    where $\Delta z = z_j - 0$ and $\Delta n_{\rm Q} \approx \mathcal{Q}dz/dt \Delta z$.
    \item Invert the cumulative star-forming SMF to solve for the progenitor mass at $z_j$, i.e., find the $m_{\star,j}^{i}$ such that $n_{\rm SF}(>m_{\star,j}^{i},z_j)$ matches the abundance found with Eq.~(\ref{eq:new_abund}).
    \item Compute the average star formation rate by taking the difference of the mass at the two timesteps $\langle\dot{m}_\star\rangle(m_\star^{i, \rm avg}, z_j) = (m_{\star,j}^{i} - m_{\star,0}^{i})/\Delta z \times dz/dt$, for $m_{\star}^{i, \rm avg} = (m_{\star,j}^{i} + m_{\star,0}^{i})/2$.
    \item Translate this into the intrinsic star formation rate by subtracting the merger driven growth contribution (Eq.~(\ref{eq:mdot_merge})) and dividing out the fraction of stellar mass returned to the ISM: $\dot{m}_\star^{\rm SFMS} = (\langle\dot{m}_\star\rangle - \dot{m}_\star^{\rm merg})/(1-R)$.
    \item Iteratively repeat steps 2-5 for the remaining redshifts $z_{j+1}$ and record the star formation rates and associated stellar masses.
\end{enumerate}
Ultimately, this results in a grid of SFRs and stellar masses associated with each of the redshift steps $z_j$ and initial masses $m_\star^i$ from which a parameterization of the star-forming MS can be inferred. In practice, we select the redshift steps $z_j$ to represent the central redshifts for which the SMF is directly measured, though we find that the results are insensitive to the resolution of this spacing given our continuous fits (see below).

\section{Applying this procedure to the observed stellar mass function}\label{sec:obs_inference}

\begin{deluxetable*}{cccccc}
\tabletypesize{\scriptsize}
\tablecaption{Summary of Stellar Mass Functions compiled for this work\label{tab:SMF}}
\tablewidth{0pt}
\tablehead{
\colhead{Source} & 
\colhead{Survey} & 
\colhead{Area (${\rm deg}^2$)} &
\colhead{SF vs. Q criterion} & 
\colhead{Stellar Mass Method} & 
\colhead{$z$ range}
}
\startdata
\cite{moustakas_primus_2013} & SDSS-\textit{GALEX} & 2504 & Rest $UVJ$ & \texttt{iSEDfit} & $z\sim 0.1$ \\
\cite{xu_pac_2025} & DESI BGS + DECaLS & 5349 & Color (Blue/Red)\tablenotemark{a} & \texttt{CIGALE}/PAC Method\tablenotemark{b}  & $z \sim 0.1$ \\
\cite{stefanon_galaxy_2021} & GREATS + S-CANDELS & 0.2 & Total SMF\tablenotemark{c} & \texttt{FAST} & $6 < z \le 10$ \\
\cite{weaver_cosmos2020_2023} & COSMOS2020 & 1.279 &  Rest $NUVrJ$ & \texttt{LePhare} & $0.2 < z \le 7.5$ \\
\cite{shuntov_stellar_2026} & COSMOS-Web & 0.42 & Rest $NUVrJ$ & \texttt{LePhare} & $0.2 < z \le 5.5$ \\
\cite{shuntov_cosmos-web_2025} & COSMOS-Web & 0.42 & Total SMF\tablenotemark{c} & \texttt{LePhare} & $6 < z \le 10.$
\enddata
\tablecomments{Summary of the survey names, star-forming vs. quiescent classification metrics, stellar mass pipelines, and redshift slices used from our compilation of literature measurements of the SMF. $UVJ$ refers to rest-frame $(U-V)$ vs. $(V-J)$ colors, while $NUVrJ$ represents rest-frame $(NUV-r)$ vs. $(r-J)$ colors. We convert all the measured stellar masses to a Chabrier IMF \citep{Chabrier03} for consistency.}
\tablenotetext{a}{Because \cite{xu_pac_2025} separate their sample into a blue/red with a $g-r$ color cut alone, we restrict their data to the $10^7-10^9 M_\odot$ subsample. Above $m_\star\gtrsim 10^{10} M_\odot$, dusty star-forming galaxies may be misclassified in the `red' subsample and thus the massive end of their inferred star-forming SMF may be incomplete \citep{ilbert_mass_2013, davidzon_cosmos2015_2017, weaver_cosmos2020_2023, shuntov_stellar_2026}.}
\tablenotetext{b}{\cite{xu_pac_2025} leverage an updated version of the Photometric objects Around Cosmic webs (PAC) method which correlates the angular positions of sources in wide photometric surveys with known spectroscopic tracers to infer their physical properties.}
\tablenotetext{c}{\cite{stefanon_galaxy_2021} and \cite{shuntov_cosmos-web_2025} measure the \textit{total} galaxy stellar mass function. Therefore, we restrict our use of those measurements to $z\gtrsim 6$, when the quiescent population is expected to comprise at most a small fraction of the total SMF at all masses (see Figure~\ref{fig:SMF_fit} for example). We thus treat these measurements as representative of the star-forming SMF at those redshifts.}
\end{deluxetable*}

\begin{figure*}
    \centering
    \includegraphics[width=\linewidth]{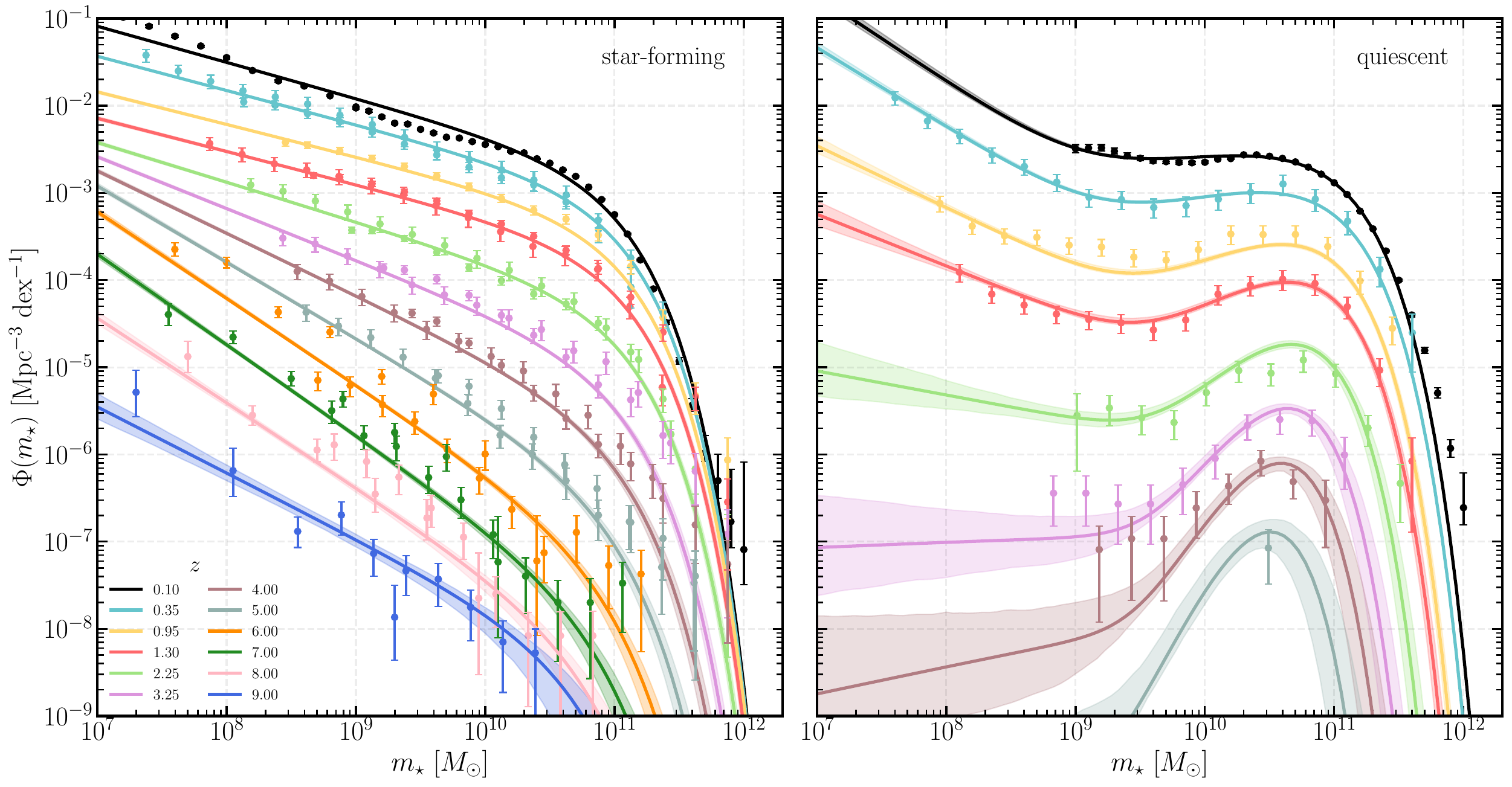}
    \caption{The \textit{intrinsic} best-fit star-forming (left) and quiescent (right) SMFs inferred from the compilation of data summarized in Table~\ref{tab:SMF} (solid curves; i.e., Eqs.~(\ref{eq:dbl_schechter}), (\ref{eq:SMF_uncertainty}), (\ref{eq:polynom_params}) with the parameters listed in Table~\ref{tab:SMF_fit_params}) for a range of redshifts (colors) offset from each other by $-0.25$ dex for clarity. The shaded colored region associated with each line represents the 16th-84th percentile uncertainties in the joint posterior distribution of the parameters, demonstrating that the uncertainty in the inferred SMF fit increases towards higher redshift and stellar mass. Note that the ostensible underestimation of the data by the solid curves at the massive end of each redshift bin is expected given that our procedure fits for the \textit{observed} SMF (Eq.~(\ref{eq:SMF_uncertainty})) whereas here we show the \textit{intrinsic} SMF.}
    \label{fig:SMF_fit}
\end{figure*}

\subsection{The observed stellar mass function}\label{sec:obs_SMF}
Using the procedure outlined in Section~\ref{sec:SMF_evolution}, we analyze a compilation of observed stellar mass function measurements for star-forming and quiescent galaxies drawn from both ground- and space-based surveys spanning $z\sim 0-9$. Some salient details of the surveys are summarized in Table~\ref{tab:SMF}. We restrict our analysis to $z\lesssim 9$, beyond which stellar mass estimates become increasingly uncertain and the rapidly declining abundance of galaxies substantially reduces the validity of our procedure.

The star-forming and quiescent stellar mass functions can, in general, be well-represented by a single and double Schechter function, respectively \citep{schechter_analytic_1976, cole_2dF_2001}.\footnote{A double power-law (DPL) functional form is sometimes used to describe the shape of the UVLF at high redshifts (see e.g., \citealp{finkelstein_coevolution_2022, finkelstein_complete_2024, donnan_jwst_2024}, and the references therein) and a similar form may be appropriate for the high-$z$, star-forming SMF \citep{shuntov_cosmos-web_2025}. However, because we limit our analysis to the range of stellar masses covered by the data, we expect that differences in the adopted functional form would minimally affect the resulting cumulative number densities and thus the inferred SFRs.} However, because of degeneracies between the Schechter parameters and the limited mass coverage of the observed SMF at high redshifts, fits to the observed mass function at individual epochs can yield mass functions that do not evolve continuously with time. When interpolated to solve Eq.~(\ref{eq:cum_sf_continuity_full}), for example, this can result in unphysical number density variations that confuse inference of the star-forming MS \citep{leja_reconciling_2015}. Thus, we enforce continuity in the evolution of the galaxy population by imposing a polynomial redshift dependence to the Schechter parameters and fit all our compiled redshift bins simultaneously.

The (logarithmic) double Schechter function is given by
\begin{equation}
    \label{eq:dbl_schechter}
    \begin{split}
    \Phi(\M)d\M &= \ln(10)\exp\big(-10^{\M-\M^*}\big)10^{\M-\M^*} \\
    &\times \Big[\phi_1^*10^{(\M-\M^*)\alpha_1} + \phi_2^*10^{(\M-\M^*)\alpha_2}\Big]\ ,
    \end{split}
\end{equation}
where $\M$ is again our logarithmic mass variable and the Schechter parameters $\M^*,\alpha_1^*, \alpha_2^*, \phi_1^*, \phi_2^*$ all depend on redshift. If $\phi_2^\star = 0$, this reduces to the single Schechter function that we use for the star-forming galaxy population. We further parameterize each of these parameters with a polynomial function and describe our fitting procedure in detail in Appendix~\ref{app:SMF_fit}.

The stellar masses inferred for the populations represented in the SMFs summarized in Table~\ref{tab:SMF} are subject to uncertainties (e.g., in the spectroscopic calibrations or SED fitting algorithms). Such uncertainties can bias the measurement of the mass function as incorrectly-characterized low mass galaxies will scatter into more massive bins and preferentially inflate the measured number densities at the high-mass end of the SMF (the so-called Eddington bias, \citealp{eddington_on_1913}). As a result, the observed mass function is a convolution of the intrinsic double Schechter (Eq.~(\ref{eq:dbl_schechter})) with an uncertainty kernel, which is generally taken to be a (log)normal distribution\footnote{Though see also \cite{ilbert_mass_2013, davidzon_cosmos2015_2017} for alternative examples, such as the product of a Gaussian and Lorentzian distribution.} \citep[e.g.,][]{behroozi_comprehensive_2010, behroozi_average_2013, behroozi_universemachine_2019, rodriguez-puebla_matching_2025}:
\begin{equation}
    \label{eq:SMF_uncertainty}
    \Phi_{\rm obs}(\M,z) = \int_{-\infty}^{\infty}\mathcal{P}(\M'-\M|\sigma_{\M}(z))\Phi(\M',z)d\M'\ ,
\end{equation}
where $\Phi$ is given by Eq.~(\ref{eq:dbl_schechter}) and $\mathcal{P}(\M'-\M|\sigma_{\M})$ is a normal distribution centered at 0 with a variance given by the fit to a compilation of stellar mass uncertainty estimates from the literature (Eqs.~(4)-(5) in \citealp{rodriguez-puebla_matching_2025}).

We fit this functional form to the measured quiescent and star-forming SMFs summarized in Table~\ref{tab:SMF} (setting the appropriate parameters to be 0 in Eq.~(\ref{eq:polynom_params}) for each population). In practice, we run a Markov Chain Monte Carlo (MCMC; \citealp{foreman-mackey_emcee_2013}) assuming a Gaussian likelihood\footnote{For asymmetric uncertainties in the SMF measurements we choose the larger of the uncertainties for our analysis.} with uniform priors on the parameters as summarized in Appendix~\ref{app:SMF_fit}. The best-fit parameters are then the maximum a posteriori estimates that result from this analysis and are reported (along with their uncertainties) in Table~\ref{tab:SMF_fit_params}. The total stellar mass function is then given by
\begin{equation}
    \label{eq:SMF_tot}
    \Phi_{\rm tot}(\M, z) = \Phi_{\rm SF}(\M, z) + \Phi_{\rm Q}(\M. z)\ ,
\end{equation}
where the two terms correspond to the star-forming and quiescent contributions to the galaxy population, respectively.
 
The results of this procedure are shown in Figure~\ref{fig:SMF_fit}. From this Figure, it is evident that the uncertainty in the SMF increases at high masses ($m_\star \gtrsim 10^{10}-10^{11} M_\odot$) and redshifts ($z\gtrsim 5$). This effect is heightened in the quiescent stellar mass function, for which the observed mass coverage grows sparse beyond $z\sim 2.5$. In turn, we note that there is systematic uncertainty between measurements from different surveys, reflecting uncertainties in stellar mass estimation, survey volume characterization, and cosmic variance quantification, for example, thought we do not attempt to account for them in this analysis beyond what is quoted in each survey. This underscores the need for wider field surveys at high redshift to reconcile these differences. 

\begin{figure*}
    \centering
    \includegraphics[width=\linewidth]{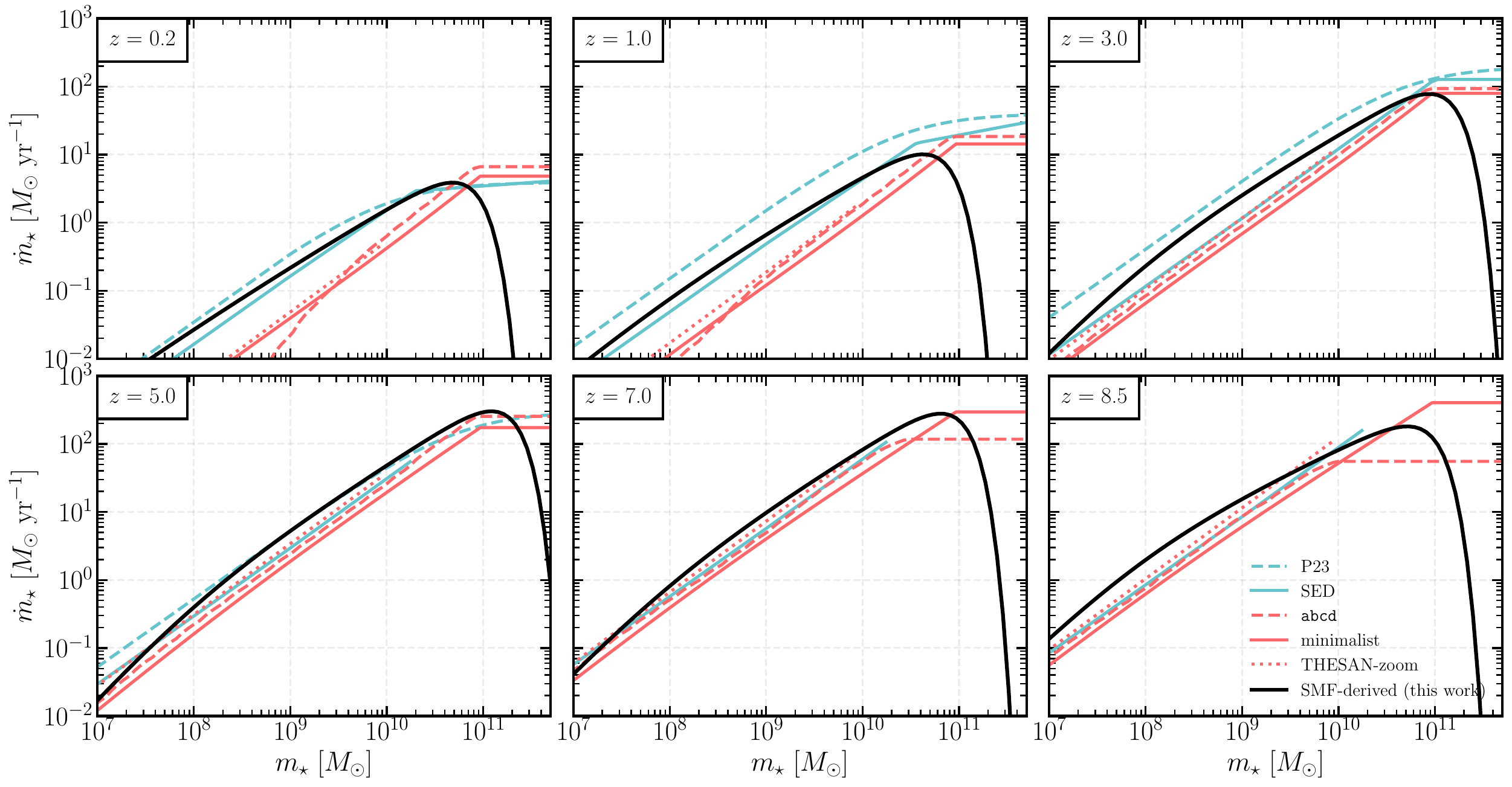}
    \caption{\textbf{The SMF-based star-forming MS broadly agrees with the star-forming MS inferred from SED fitting procedures with non-parametric star formation histories, albeit with a steepening slope at low masses.} The full star-forming MS for a subset of the theoretical (red) and observational (blue) results shown in Figure~\ref{fig:sSFR} at different redshifts (different panels). The black curve corresponds to the star-forming MS inferred from the redshift evolution of the SMF, the key result of this work.}
    \label{fig:MS_comparisons}
\end{figure*}

\subsection{The SMF star-forming main sequence}\label{sec:SMF_MS}
With this in hand, we consider a star-forming main sequence constructed directly from the evolution of the observed SMF as outlined in Section~\ref{sec:continuity}. Briefly, we infer the SFRs necessary to preserve the redshift evolution of the star-forming stellar mass function, accounting for the non-conservative effects of mergers and quenching, which will modify the overall abundance of star-forming galaxies. By solving Eq.~(\ref{eq:cum_sf_continuity_full}) with the continuous SMF described in Section~\ref{sec:obs_SMF}, we infer a grid of mass growth rates and associated stellar masses at the discrete redshift bins for which the observed SMFs are reported. We subtract the mass growth due to mergers and divide by the return fraction of material to compute the mean SFR. 
We find that a somewhat similar function to Eq.~(\ref{eq:p23_fit}) fits this grid of star-forming MS measurements well, albeit with some key differences:
\begin{equation}
    \label{eq:MS_fit_dpl}
    \dot{m}_\star = \frac{\dot{m}_\star^{\rm norm} m_{\star,8}^{\beta}}{1 + m_{\star,8}^{\beta-\alpha}} 10^{(m_\star/m_{\rm turn})^2}\ ,
\end{equation}
where $m_{\star,8} = m_\star/10^8 M_\odot$. Here, we have four free parameters that evolve with redshift: the SFR normalization $\dot{m}_\star^{\rm norm}$, the low-mass slope $\beta$, the high-mass slope $\alpha$, and the turnover mass $m_{\rm turn}$. Thus, our star-forming main sequence derived from the observed SMF is characterized by a double power law over most masses and a turnover to an exponential cutoff at the high mass end ($m_{\rm turn}\sim 10^{10}-10^{11} M_\odot$), with the latter feature driven by the overwhelming contribution of halo mergers to mass growth in the most massive systems. We summarize the redshift evolution of each parameter and the fitting procedure in Appendix~\ref{app:star-forming MS_fit} and list the best fit parameters in Table~\ref{tab:star-forming MS_fit_params}.

We display the resulting fit in Figures~\ref{fig:sSFR} and \ref{fig:MS_comparisons} as a solid black line along with a shaded region representing the uncertainties propagated from the joint posterior of the star-forming and quiescent SMF fitting procedure.

\section{Discussion}\label{sec:discussion}

\subsection{The normalization and evolution of the SFMS}\label{sec:star-forming MS_evolution}
We begin by exploring the redshift evolution of the MS through Figure~\ref{fig:sSFR}, which shows the time evolution of the sSFR at a fixed stellar mass of $m_\star\sim 10^{9.5} M_\odot$.

First, we recover the longstanding discrepancy in which theoretical models underpredict SFRs at fixed stellar mass, though the tension with the SMF-derived MS is substantially weaker than some previous estimates (\citealp{speagle_highly_2014} and \citealp{popesso_main_2023}). As discussed in \cite{leja_new_2022}, the revised stellar masses and star formation rates derived from the use of non-parametric SFHs in SED fitting routines can reconcile some of the differences between theoretical and observational results, evidenced by the agreement between the \cite{leja_new_2022} and IllustrisTNG MS.\footnote{Some authors have additionally noted an issue in the dust law and resulting calibration used to infer SFRs from H$\alpha$ luminosities in low-metallicity environments \citep{shapley_JWST_2023}. While this modification would introduce a $-0.34$ dex offset to the SFR measured for low-metallicity systems in \cite{speagle_highly_2014} and \cite{popesso_main_2023} (which would be most pronounced at $z\gtrsim 3$) and bring the normalization of the observed star-forming MS measurements closer in agreement, it would not reconcile the markedly different redshift evolution discussed here as well.}

Next, we find that, in their overlap region ($z\sim 2-3$), the JWST/NIRSpec-based star-forming MS measurements agree very well with the pre-JWST inference carried out with \texttt{Prospector} \citep{leja_new_2022}, and suggest a distinct redshift evolution from the \cite{speagle_highly_2014} and \cite{popesso_main_2023} compilations. In turn, the normalization of the star-forming MS inferred from JWST/NIRCam photometric samples is lower than seen in the spectroscopic sample, though the broad redshift evolution is similar \citep{simmonds_bursting_2025}. Comparing the SMF-derived star-forming MS to these samples, we find a similarly gentle redshift evolution at $z\lesssim 2$ (compared with steep decline seen in the main sequence compilations) and a strong agreement with the \cite{leja_new_2022} measurements derived from pre-JWST samples.\footnote{Though not shown, the MS inferred from stacking of sources observed in the far-infrared (FIR) with the \textit{Herschel} and \textit{James Clerk Maxwell} Telescopes also displays a similar (lower) normalization and gentle redshift evolution to the MS presented in \cite{leja_new_2022} and derived in this work \citep{koprowski_charting_2024}.} At higher redshifts, the SMF-derived star-forming MS agrees exceptionally well with the JWST/NIRSpec results \citep{clarke_star_2025}, and suggests an increasing trend of SFRs to the earliest times.

Focusing now on the theoretical results shown in Figure~\ref{fig:sSFR}, we find broad agreement between most models in the literature, especially in the redshift evolution at fixed stellar mass, suggesting that \textit{DM halo accretion rates are the fundamental --- and ubiquitous --- driver of the long-term predicted star-forming MS evolution}, as has been argued in previous work \citep{dave_analytic_2012, dekel_toy_2013}. We emphasize the consistency of this behavior across a range of modeling choices, most notably in the parameterization of the underlying physics and complexity in the mathematical and numerical calculations. In addition, we note that the two analytical models, the minimalist and \texttt{abcd} are only calibrated to luminosity function measurements at $z>6$, while the feedback model in the numerical simulations --- both of which rely on the same galaxy formation model --- is calibrated to reproduce SMFs at $z\sim 0$ \citep{pillepich_simulating_2018}.

A well-established feature of theoretical models is an underprediction of most pre-JWST measurements of the MS, across different physical assumptions. The primary discrepancy between different theoretical models is in their relative normalization, which is in part a result of the treatment of stellar feedback and stochasticity in the models. That is, the inclusion of bursty star formation in $\texttt{abcd}$ compared to the minimalist model introduces higher mean SFRs at fixed stellar mass (induced by stochasticity in the SFHs), thereby offsetting the curves while maintaining the long-term redshift evolution. In kind, the THESAN-zoom simulation, which includes a more detailed treatment of burstiness resulting from gas cycling and mergers, finds a remarkably similar redshift evolution and offset to the vastly simpler \texttt{abcd} model. Though the overlap is small, the IllustrisTNG simulated star-forming MS also traces a similar redshift evolution, albeit with a slightly higher ($<$2x) normalization. Notably, the inferred sSFR from the SMF evolution follows a similar trajectory to the theoretical literature, reinforcing the conclusion that halo accretion rates drive the long term evolution of galaxy SFRs.

Because the star-forming MS inferred here has a much gentler redshift evolution than previous concordance compilations of the MS, we emphasize that analyses that relied on these measurements may need to be revisited. In a recent example, \cite{kimmig_built_2026} explore the implications of evolution along and around the \cite{popesso_main_2023} star-forming main sequence on the $z\sim 0$ observed galaxy population. Our star-forming MS, which implies higher rates of star formation at early times and lower rates at $z\lesssim 5$ (compared with the \citealp{popesso_main_2023} star-forming MS at fixed stellar mass), would then yield qualitatively different age distributions and imply different star formation histories for the local galaxy population. We defer a detailed exploration of such implications to future work, and instead here simply caution against applying star-forming main sequence measurements from the literature without appropriately characterizing the systematic and modeling uncertainties that exist.

\subsection{The shape of the star-forming main sequence}\label{sec:star-forming MS_shape}
In Figure~\ref{fig:MS_comparisons}, we show the full SFR-stellar mass relation and focus our discussion on the inferred shape.

Across a wide range of redshifts, we find that most measurements of the star-forming MS show a similar low-mass slope and turn over at $m_\star\gtrsim 10^{10.5} M_\odot$, when quenching and mergers become significant. Below that mass, the star-forming MS is largely characterized by a power law, though a double power law is necessary at the highest redshifts. The slopes between $10^{8.5} < m_\star < 10^{10.5} M_\odot$ tend to be fairly constant over time ($\alpha\sim 0.75-0.8$, before declining to $\alpha\sim 0.5$ at $z\sim 9$; see Appendix~\ref{app:star-forming MS_fit}). This slope agrees very well with those measured in \cite{speagle_highly_2014} and \cite{clarke_star_2025}, both of which are limited to galaxy samples above $m_\star \geq 10^9 M_\odot$ and $10^{8.5} M_\odot$, respectively. While the normalization of the SMF-based star-forming MS is closer to the SED fitting results, we note that \cite{leja_new_2022} find a slightly steeper star-forming MS ($\alpha\sim 1$), consistent with \cite{popesso_main_2023}.

At the lowest masses, the slope is comparatively steeper, and grows from unity at $z\sim 0$ to $\beta\sim 1.5$ at $z\sim 9$. However, there are no directly observed star-forming MS measurements that probe these low stellar masses, so we cannot compare this result with existing work. We note that this is an example of the power of inferring the star-forming MS from the SMF --- because mass completeness is already folded into the SMF measurements, such a star-forming main sequence is well-characterized over any range of masses for which SMF data is available (and is thus less sensitive to biases that can arise in direct galaxy surveys; see the discussions in \citealp{davidzon_alternate_2018, clarke_star_2025}, for example).

Over most masses and redshifts, the SMF-derived star-forming MS shows a similar slope to that of the minimalist and \texttt{abcd} models, both of which are characterized by energy-regulated stellar feedback. We note that the simulated star-forming MS from THESAN-zoom shows a steeper slope (comparable to the $\alpha\sim 1$ found in \citealp{popesso_main_2023}). However, this star-forming MS is derived over a very different mass range than the observed literature (the majority of the THESAN-zoom systems fall between $m_\star \sim 10^7-10^9 M_\odot$, while most observed MS estimates are mass-complete above $m_\star\gtrsim 10^9 M_\odot$), so it may be more appropriate to compare with the steeper low mass slope which we find here. This varying slope of the star-forming MS with galaxy mass and redshift at the lowest masses has not been seen in previous work, and is potentially indicative of different feedback and environmental mechanisms at play in these systems, though we defer a detailed exploration of this effect and comparison with theoretical models to future work.

\subsection{Star formation efficiencies}\label{sec:SFE}
\begin{figure}
    \centering
    \includegraphics[width=\linewidth]{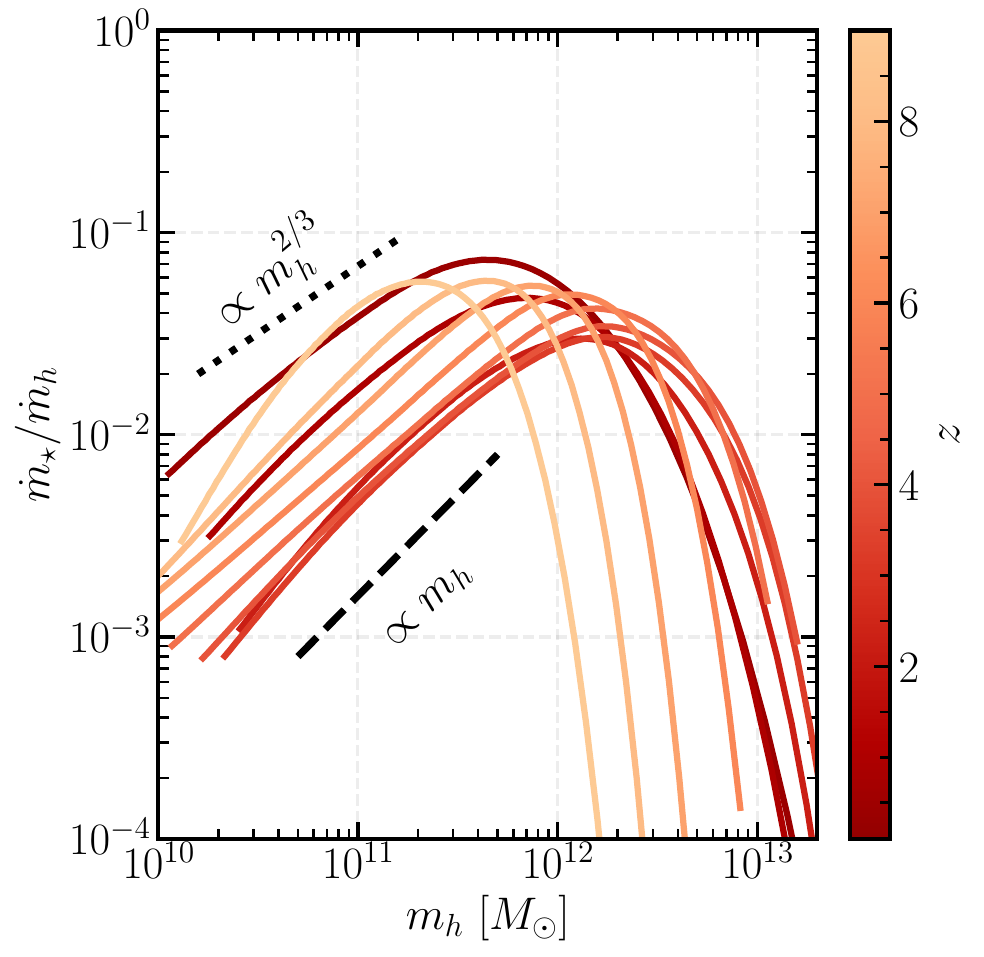}
    \caption{\textbf{The instantaneous star formation efficiency inferred from the SMF-based star-forming MS suggests both an evolution of the low mass slope and peak mass.} The ratio of star formation rate associated with the SMF-derived star-forming MS (Section~\ref{sec:SMF_MS}) to DM halo accretion rates from the Millennium simulations \citep{fakhouri_merger2010} at a range of redshifts (colors), shown as a function of halo mass. We associate galaxies with host halos by abundance matching the SMF and HMF (Eq.~(\ref{eq:SHMR})). We also show power laws with slopes of 1 and 2/3 in dashed and dotted black lines, respectively, to help guide the eye.}
    \label{fig:SFE}
\end{figure}

We next extend our analysis of the star-forming MS by considering the implications for the galaxy-halo connection. That is, using a similar abundance matching procedure to associate galaxies with halos (connecting the total SMF with the halo mass function, see Eq.~(\ref{eq:SHMR})), we associate galaxy and halo growth rates, where $\dot{m}_\star$ is our inferred star-forming MS and $\dot{m_h}$ is the fit to the mean growth rate measured from the Millennium DM growth simulation \citep{fakhouri_merger2010}. Carrying out this procedure at a range of redshifts yields the star formation efficiency (SFE)\footnote{Here we focus on the \textit{instantaneous} SFE, defined as the ratio of the SFR to halo mass accretion rate, though we note that other works use a similar term for the integrated SFE, or the ratio of the stellar to halo mass (which we call the SHMR, $f_\star$).} evolution shown in Figure~\ref{fig:SFE}.

The low mass slope of the SFE is a result of the scaling of stellar feedback with halo mass, which itself provides a lens into the feedback mechanisms at play \citep{sun_constraints_2016, furlanetto_minimalist_2017, mirocha_prospects_2020}. A slope of $\propto m_h^{2/3}$ ($\propto m_h^{1/3}$) is generally associated with energy (momentum) regulated feedback. We find that the SFE associated with our inferred star-forming MS and compiled SMF has a low mass slope that gradually flattens over time. That is, if the SFE $\propto m_h^{\gamma}$ at low masses, we find that $\gamma \gtrsim 1$ at $z\gtrsim 7$, flattens to $\gamma \sim 1$ at intermediate times, and approaches $\gamma\sim 2/3$ at the lowest redshifts. 

The peak and high mass slope of the SFE evolves as a result of galaxy quenching mechanisms \citep{behroozi_average_2013, behroozi_universemachine_2019}. We find that the peak mass of the SFE grows from a few $\times 10^{11} M_\odot$ to $\sim 10^{12} M_\odot$ at $z\sim 2-3$, before falling to $m_h^{\rm peak}\sim 3\times 10^{11} M_\odot$ at $z\sim 0$. However, diagnosing the physical origin of these features requires careful theoretical modeling which is beyond the scope of this work.

\subsection{The global star formation rate density}\label{sec:SFRD}
An observable that \textit{combines} the evolution of the SMF and star-forming MS is the global star formation rate density. Inferred from the UV luminosity density (which itself is an integrated statistic of the UVLF), the SFRD can be less sensitive to completeness issues and other systematics compared with the SMF (as robustly characterizing the mass completeness of a survey can be challenging given the often nontrivial relationship between galaxy mass and light).\footnote{We do, however, note that inferring the SFRD from the UV and IR luminosity density requires assumptions for the dust correction (in the former case) and the mapping between UV/IR luminosities and SFRs (in both cases), which can introduce systematic uncertainties in this measurement as well.} In this sense, the integrated evolution of SFRs across time provides a somewhat independent lens into the consistency of the scaling relations considered here with the full galaxy population.

Given our evolving SMF and the star-forming MS variations, the SFRD can be estimated as
\begin{equation}
    \label{eq:SFRD}
    \rho_{\rm SFR} = \int_{\M_{\rm min}}^{\M_{\rm max}}  \dot{m}_\star^{\rm SFMS} \Phi_{\rm SF} d\M\ ,
\end{equation}
where the lower integration bound is set by the observing limits of the UVLFs used to compute the luminosity density and the upper bound is taken to be $\M_{\rm max} = 12.5$. To compare against the \cite{madau_cosmic_2014} inference of the SFRD evolution, we enforce a lower integration bound corresponding to their limiting faint-end luminosity of $L_{\rm min} = 0.03 L_\star \approx 0.03 \times 10^{10} L_\odot$. Applying the $L_{\rm UV}-$SFR conversion factor quoted in \cite{madau_cosmic_2014}, $\kappa_{\rm UV} = 1.15\times 10^{-28}\ {\rm M_\odot\ yr^{-1}\ (erg\ s^{-1}\ Hz^{-1})^{-1}}$ at a wavelength of $\lambda_{\rm UV} = 1500~\text{\AA}$, we have $\dot{m}_{\rm \star, min} = \kappa_{\rm UV}L_{\rm min}/\nu_{\rm UV} \approx 6.61\times 10^{-2}\ M_\odot\ {\rm yr^{-1}}$. The lower bound of our integration at each redshift is then $\M_{\rm min} = \log_{10}(m(\dot{m}_{\rm \star,min}))$, evaluated by inverting the star-forming main sequences listed above.

\begin{figure}
    \centering
    \includegraphics[width=\linewidth]{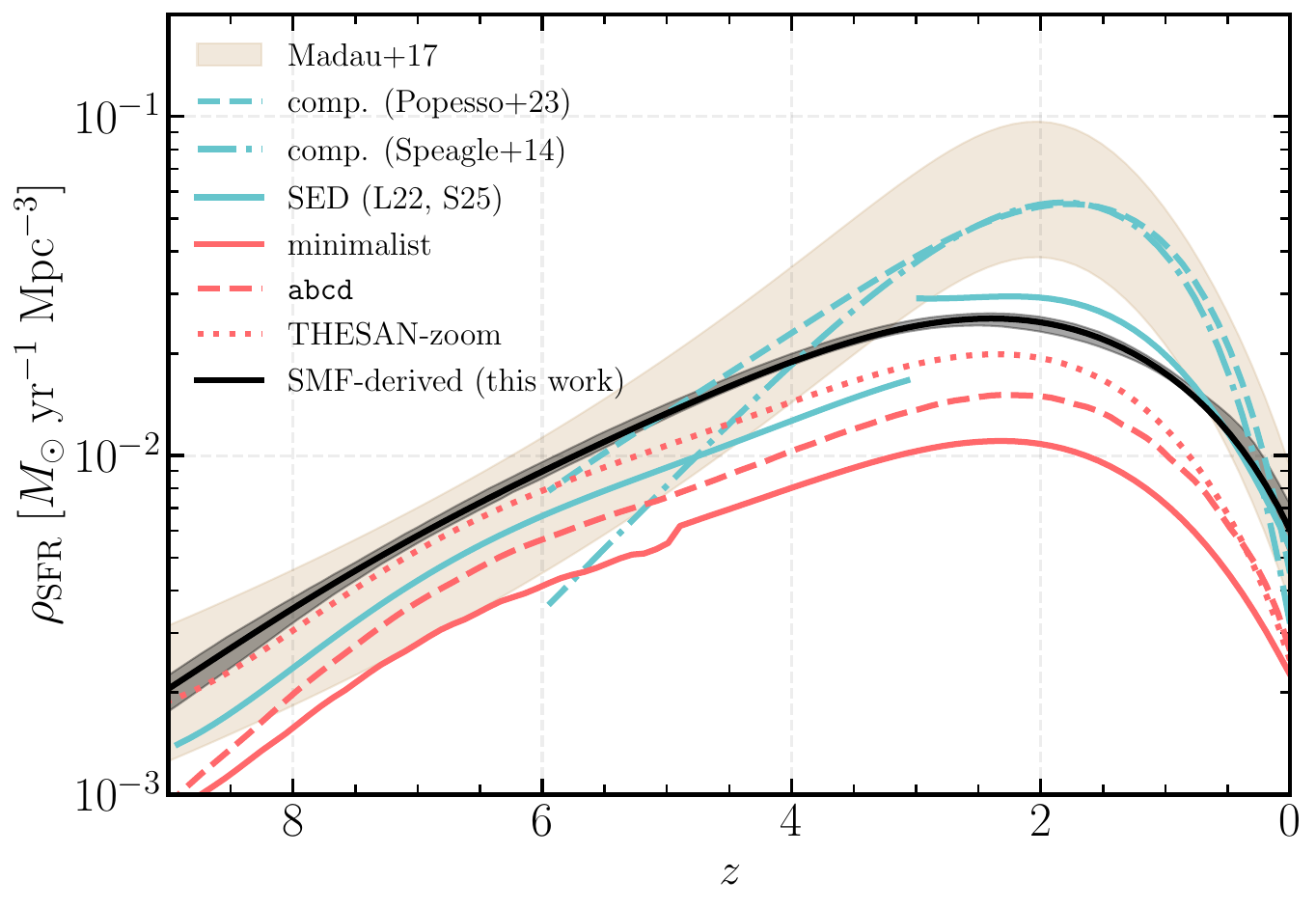}
    \caption{\textbf{The SFRD predicted by the SMF-derived main sequence is broadly consistent with independent measurements \citep{madau_radiation_2017}, though the peak of cosmic star formation is smaller by a factor of a few at cosmic noon.} The SFRD inferred by integrating the star-forming MS measurements described in Section~\ref{sec:star-forming MS_overview} over the SMF compilation used in this work (Section~\ref{sec:obs_SMF}), compared with the integrated \cite{madau_radiation_2017} SFRD based on a compilation of galaxy luminosity density measurements (tan shaded region). We again show theoretical predictions in red and observed measurements in blue with the same linestyles as Figures~\ref{fig:sSFR} and \ref{fig:MS_comparisons}. The SMF-derived main sequence is shown as a black solid line with a gray shaded region highlighting the uncertainty associated with the SMF fitting procedure.}
    \label{fig:SFRD}
\end{figure}

We show the results of this integration in Figure~\ref{fig:SFRD}. While we find a qualitatively similar redshift evolution to the global SFRD inferred from the galaxy luminosity density \citep{madau_radiation_2017}, we find that the peak of cosmic star formation is somewhat suppressed in our results. That is, because the redshift evolution of the SMF-based star-forming MS summarized in Figure~\ref{fig:sSFR} is gentler than preceding compilations, the $z\sim 2$ peak in the SFRD is smaller by a factor of $\sim 2-3$. This is a generic feature of all the star-forming main sequences that evolve in a similar manner to DM halo accretion rates. Indeed, the star-forming MS measurements with a higher normalization and steeper decline to low redshifts (i.e., the \citealp{speagle_highly_2014} and \citealp{popesso_main_2023} compilations) reproduce the peak of cosmic star formation much more closely. Investigating this discrepancy in detail is beyond the scope of this work and we reserve this investigation for future work.\footnote{See \cite{fu_how_2026} for a related empirical investigation of the SFEs and associated SFHs implied by high-redshift luminosity functions and the SFRD.} 

\section{Conclusion}\label{sec:conclusion}
In this work, we have carried out a statistical inference of the star-forming main sequence from the redshift evolution of the stellar mass function. In particular, we construct a continuous fit to a compilation of the latest observed SMF measurements from $z=0.1-9$, combining both ground- and space-based samples, including results from JWST. We quantify the effects of mergers and galaxy quenching on modifying the observed SMF and use a continuity equation for galaxy mass growth to infer the star formation history of the universe out to the first 500 Myr.

Our conclusions can be summarized as follows:
\begin{enumerate}
    \item The star-forming MS required by the redshift evolution of the SMF agrees very well with JWST-based spectroscopic surveys of star-forming galaxies from $z\sim 2-7$ across a broad range of masses. This star-forming MS is also in good agreement with SFRs inferred from observations of galaxies at $z\lesssim 3$ studied with non-parametric SFHs and SED modeling. The normalization of this main sequence is lower than compilations of other measurements from the literature at fixed stellar mass, though the two converge to the same local value. 
    \item The agreement with the most sensitive JWST spectroscopy demonstrates the power of our procedure to characterize the galaxy population down to lower masses than conventional surveys, which target individual sources and can suffer from completeness challenges at the faint end as a result. We thus advocate for continuing such phenomenological analyses as SMF measurements extend to the faint galaxy population ($m_\star \lesssim 10^8 M_\odot$) as a complementary tool to the more costly spectroscopic surveys that may not be able to statistically probe this regime.
    \item The SMF-based star-forming MS aligns much more closely with predictions of theoretical models based on hierarchical structure formation and DM halo growth than most previous star-forming MS estimates, indicating that improved constraints on stellar masses measured with JWST could be alleviating this tension. In turn, the redshift evolution of the SMF-derived star-forming MS more closely mirrors the redshift evolution of DM halo accretion rates, suggesting that halo growth is indeed the principal driver of the long term evolution of mean galaxy star formation histories.
    \item The slope of the star-forming MS is fairly constant across time over most masses, with a slight flattening at high redshifts ($\alpha\sim 0.75$ at $z=0$ to $\alpha\sim 0.5$ at $z\sim 9$). At the lowest masses ($m_\star \lesssim 10^{8.5} M_\odot$), the slope of the star-forming MS steepens to $\beta\sim 1.5$, suggesting that different feedback mechanisms may be regulating the shape in different galaxy mass regimes. 
    \item The gentler redshift evolution and similarity of the SMF-based star-forming MS with DM halo accretion rates reinforce the picture that DM halo growth may be the dominant driver of the long term evolution of galaxy star formation rates. However, we find evidence that the halo-scale SFE evolves with time, declining and then rising for halos of fixed mass between $z\sim 9$ and 0.
    \item The integral of the star-forming MS and SMF explored in this work yield a SFRD that does not reproduce the strong peak of cosmic star formation seen in studies of the UV luminosity density, reproducing the longstanding tension between stellar mass density and SFRD seen in the literature.
\end{enumerate}
The distinct shape and redshift evolution of the star-forming MS derived in this work relative to suggest that the literature is still not converged, and corrections for known systematic uncertainties in compilations of main sequence measurements may not reveal the \textit{true} mean SFRs of the galaxy population. In addition, while the SMF-derived MS circumvents some systematics present in other MS measurement techniques, there remain uncertainties in e.g., the underlying stellar mass estimates, the evolving stellar IMF, and survey corrections for cosmic variance, which can all bias the results of this analysis. Subsequent work with wider field surveys and comparative analyses across modeling assumptions will be necessary to tighten these uncertainties and refine estimates of the SMF-derived star-forming MS. With these uncertainties in mind, studies that rest on the star-forming MS as a fundamental input characterizing galaxy demographics across time should exercise caution and test for sensitivity to these uncertainties. While the closer agreement with theoretical literature noted here is perhaps comforting, there are nevertheless features in the star-forming MS and its time evolution that beget further exploration. Thus, it is evident that our understanding of the star formation history of the universe remains incomplete, though now with JWST we are moving ever closer to a cohesive observational picture of galaxy populations across time.

\section*{Acknowledgments}\label{sec:acknowledgements}
S.H. thanks Leonardo Clarke, Olivia Cooper, and Greg Bryan for insightful conversations. S.H. is supported by the National Science Foundation Graduate Research Fellowship Program under Grant No. DGE-2034835. Any opinions, findings, and conclusions or recommendations expressed in this material are those of the authors and do not necessarily reflect the views of the National Science Foundation. S.H. acknowledges support from the Future Investigators in NASA Earth and Space Science and Technology (FINESST) Grant No. 80NSSC23K1432, the UCLA Michael A. Jura Fellowship, and the UCLA Consortium for Developing Leadership in Science (CDLS) Fellowship. S.H. and S.R.F. were supported by NASA through award 80NSSC22K0818 and by the National Science Foundation through award AST-2510939.  S.N. thanks the support of NASA grant No. 80NSSC24K0773 (ATP-23-ATP23-0149) and Howard and Astrid Preston for their generous support.

We acknowledge that the location where this work took place, the University of California, Los Angeles, lies on indigenous land. The Gabrielino/Tongva peoples are the traditional land caretakers of Tovaangar (the Los Angeles basin and So. Channel Islands). 

This work has made extensive use of NASA's Astrophysics Data System (\href{http://ui.adsabs.harvard.edu/}{http://ui.adsabs.harvard.edu/}) and the arXiv e-Print service (\href{http://arxiv.org}{http://arxiv.org}), as well as the following softwares: \textsc{matplotlib} \citep{Matplotlib}, \textsc{numpy} \citep{numpy}, \textsc{astropy} \citep{Astropy}, \textsc{emcee} \citep{foreman-mackey_emcee_2013}, and \textsc{scipy} \citep{Scipy}.

\section*{Data availability}\label{sec:data}
The SMF and star-forming MS fit parameters and a Python implementation of these functions is available at the following repository: \href{https://github.com/hegdesahil/SMF_SFMS}{https://github.com/hegdesahil/SMF\_SFMS}.

\appendix

\begin{figure}
    \centering
    \includegraphics[width=\linewidth]{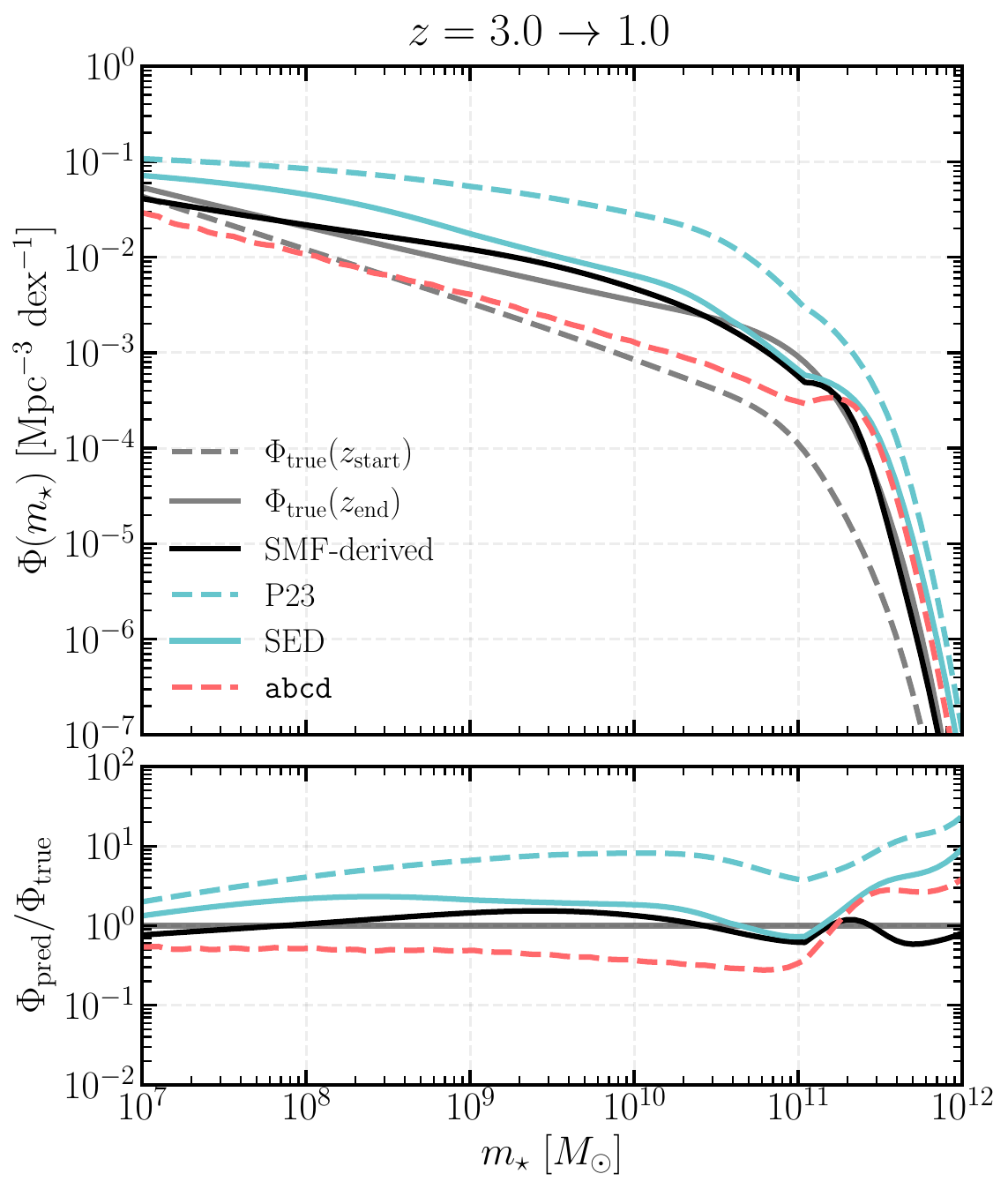}
    \caption{\textbf{Literature consensus estimates for the star-forming MS vastly overpredict the redshift evolution of the SMF, though SED fitting-based measurements perform better.} The SMF at evolved from $z=3$ to $z=1$ using various estimates for the star-forming MS to quantify the stellar mass `drift' of galaxies over time. The true SMFs at the starting and ending redshift are shown in gray (dashed and solid, respectively), and the evolved SMF using various star-forming MS estimates are shown in different colors (\citealp{popesso_main_2023} in dashed blue, \citealp{leja_new_2022} in solid blue, the \texttt{abcd} model of \citealp{hegde_efficient_2025} in dashed red, and the SMF-based star-forming MS computed in this work in solid black. The lower panel shows the ratio of these predicted SMFs to the true SMF at $z_{\rm end} = 1$.}
    \label{fig:SMF_evolution}
\end{figure}

\begin{figure}
    \centering
    \includegraphics[width=\linewidth]{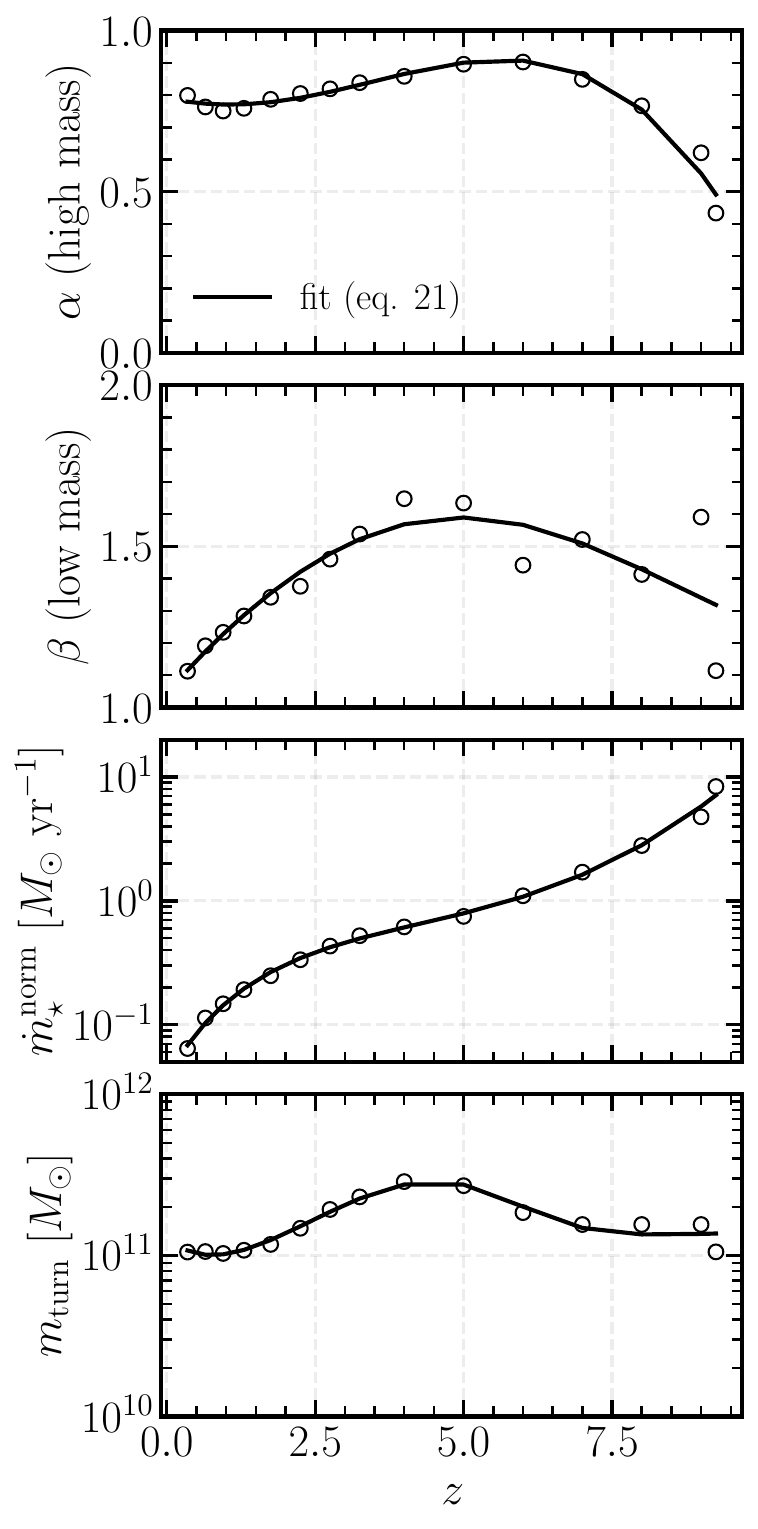}
    \caption{\textbf{The high mass slope of the star-forming MS is roughly constant to $z\sim 6$ and the low mass slope deviation becomes more pronounced beyond this point as well.} Redshift evolution of the fit parameters for the SMF-based star-forming MS (Eqs.~(\ref{eq:MS_fit_dpl}-\ref{eq:logistic}; each parameter is shown in an individual panel). The points correspond to the fit parameters computed at each redshift and the line shows our redshift-dependent fit to those parameters (Eq.~(\ref{eq:MS_fit_z_dep})).}
    \label{fig:star-forming MS_params_zDep}
\end{figure}

\section{Evolving the SMF with the SFMS}\label{app:evolving_SMF}
As a validation of our procedure and complementary demonstration of the scaling relation inconsistency, we evolve the SMF \textit{forward in time} with the star-forming MS as an input. However, because our procedure for quantifying the effects of mergers on the SMF is only calibrated to trace the progenitor evolution of the SMF (i.e., the backwards evolution), we modify Eq.~(\ref{eq:continuity}) to carry out the integration. 

In particular, we evolve
\begin{equation}\label{eq:continuity_tot}
    \frac{\partial \Phi_{\rm tot}(m_\star, t)}{\partial t} + \frac{1}{\ln10}\frac{\partial}{\partial \M}\bigg[\frac{\Phi_{\rm tot} \langle\dot{m}_\star^{\rm tot}\rangle_{\rm all}}{m_\star}\bigg] = 0,
\end{equation}
where now we are integrating the \textit{total} SMF (i.e., Eq.~(\ref{eq:SMF_tot})), rather than the star-forming SMF. In this case, the quiescent galaxies are included in the population, so the quenching term on the RHS of the expression disappears, and we quantify the nonconservative effects of mergers in the descendant evolution direction using the results of \cite{torrey_forward_2017}. \cite{torrey_forward_2017} provide a fitting function for the fraction of the total galaxy population that survives merging as a function of stellar mass, initial redshift, and final redshift (see their Appendix A). Because this fit is only provided for $z\leq 3$, we limit the analysis in this section to that range.

Because we are now evolving the full galaxy population, the stellar mass drift of the population (i.e., $\langle \dot{m}_\star^{\rm tot}\rangle_{\rm all}$) is not just the sum of the mean SFR and the mass growth due to mergers. In other words, the average SFR is given by a weighted sum of the star-forming and quiescent star-forming main sequences:
\begin{equation}\label{eq:avg_SFR}
    \begin{split}
    \dot{m}^{\rm SF}_{\star, \rm all} &= \frac{\Phi_{\rm SF}\dot{m}_\star^{\rm SFMS} + \Phi_{\rm Q}\dot{m}_\star^{\rm Q}}{\Phi_{\rm tot}}(1-R) \\&\approx  \frac{\Phi_{\rm SF} }{\Phi_{\rm tot}}\dot{m}_\star^{\rm SFMS}(1-R) \equiv f_{\rm SF}(1-R)\dot{m}_\star^{\rm SFMS},
    \end{split}
\end{equation}
where $\dot{m}_\star^{\rm SFMS}$ is the SFR implied by the star-forming MS and for the latter equivalences we have assumed that $\dot{m}_\star^{\rm q}\ll \dot{m}_\star^{\rm SFMS}$ and introduced the fraction of star-forming galaxies $f_{\rm SF}$. With this, the stellar mass velocity of the population is $\langle \dot{m}_\star^{\rm tot}\rangle_{\rm all} = \dot{m}^{\rm SF}_{\star, \rm all} + \dot{m}_\star^{\rm merg} = f_{\rm SF}(1-R)\dot{m}_\star^{\rm SFMS} + \dot{m}_\star^{\rm merg}$. As in \cite{leja_reconciling_2015}, the fraction of star forming galaxies can be inferred from the ratio of the star-forming to total stellar mass functions, and is well-represented by an 8 parameter hyperbolic tangent, capped at 1. That is,
\begin{equation}
    f_{\rm SF}(m_\star, z) = \min\bigg[(f_0(z) - f_a(z))\tanh\big[a(\M-b)\big], 1\bigg],
\end{equation}
where the normalization $f_0(z) = f_{00} + f_{01}z+f_{02}z^2$ and asymptotic values $f_a(z) = f_{a0} + f_{a1}z+f_{a2}z^2$ encode the redshift dependence.

We integrate the SMF following Eq.~(\ref{eq:continuity_tot}) from $z_{\rm start}= 3$ to $z_{\rm end} = 1$ and vary the underlying star-forming MS in Figure~\ref{fig:SMF_evolution}. The qualitative behavior expected from the comparisons made in Figures~\ref{fig:sSFR} and \ref{fig:MS_comparisons} and discussed in Sections~\ref{sec:introduction} and \ref{sec:discussion} are reflected here as well. That is, the compiled star-forming main sequence from \cite{popesso_main_2023} builds up too much stellar mass, overpredicting the $z=1$ mass function by nearly an order of magnitude at all masses. Theoretical models, represented with the \texttt{abcd} curve, perform better, but slightly underpredict the observed mass function. Of the literature star-forming MS estimates available, those inferred from nonparametric SED fitting \citep{leja_new_2022} perform best, though the star-forming MS inferred from the mass function suggests somewhat smaller star formation rates.

\begin{deluxetable}{lcccc}
\tablewidth{0pt}
\tablecaption{Maximum a posteriori estimates for Schechter parameters of the compiled SMFs (Eqs.~(\ref{eq:dbl_schechter})-(\ref{eq:SMF_tot}) and shown in Figure~\ref{fig:SMF_fit})\label{tab:SMF_fit_params}).}
\tablehead{
  \colhead{parameter ($p$)} &
  \colhead{$p_0$} &
  \colhead{$p_1$} &
  \colhead{$p_2$} &
  \colhead{$p_3$}
}
\startdata
\cutinhead{Star-forming}
$\M^*$
  & $10.8927^{+0.0054}_{-0.0053}$ & $0.049^{+0.032}_{-0.032}$ & $-0.015^{+0.016}_{-0.015}$ & $0.0006^{+0.0016}_{-0.0017}$ \\
$\log_{10}\phi^{*}_{1}$
  & $-3.0620^{+0.0067}_{-0.0071}$ & $-0.006^{+0.039}_{-0.036}$ & $-0.121^{+0.017}_{-0.018}$ & $0.0098^{+0.0018}_{-0.0017}$ \\
$\alpha_{1}$
  & $-1.4246^{+0.0028}_{-0.0027}$ & $0.117^{+0.014}_{-0.013}$ & $-0.0718^{+0.0055}_{-0.0057}$ & $0.00607^{+0.00053}_{-0.00053}$ \\
$\log_{10}\phi^{*}_{2}$ & --- & --- & --- & --- \\
$\alpha_{2}$ & --- & --- & --- & ---  \\
\cutinhead{Quiescent}
$\M^*$
  & $10.8548^{+0.0050}_{-0.0051}$ &
$-0.174^{+0.035}_{-0.033}$ & --- & --- \\
$\log_{10}\phi^{*}_{1}$
  & $-2.6255^{+0.0067}_{-0.0068}$ &
$-0.384^{+0.052}_{-0.058}$ &
$-0.050^{+0.025}_{-0.024}$ & --- \\
$\alpha_{1}$
  & $-0.765^{+0.014}_{-0.014}$ &
$0.62^{+0.11}_{-0.11}$ & --- & --- \\
$\log_{10}\phi^{*}_{2}$
  & $-4.925^{+0.088}_{-0.054}$ &
$0.11^{+0.23}_{-0.25}$&
$-0.133^{+0.067}_{-0.066}$ & --- \\
$\alpha_{2}$
  & $-2.029^{+0.038}_{-0.032}$ &
$0.335^{+0.069}_{-0.081}$ & --- & --- \\
\enddata
\tablecomments{
  The uncertainties here represent the 16th-84th percentile range of each parameter. Because we show the logarithmic Schechter function, the normalizations $\phi^*$ are given in units of $[\phi^*] = {\rm Mpc^{-3}\ dex^{-1}}$.
}
\end{deluxetable}

\begin{deluxetable*}{cccccccc}
\tabletypesize{\scriptsize}
\tablecaption{Best-fit parameters for the SMF-based star-forming MS described in Eqs.~(\ref{eq:MS_fit_dpl}), (\ref{eq:MS_fit_z_dep}), and (\ref{eq:logistic}).\label{tab:star-forming MS_fit_params}}
\tablewidth{0pt}
\tablehead{
\colhead{parameter ($r$)} &
\colhead{$l_1$} & 
\colhead{$l_2$} & 
\colhead{$l_3$} &
\colhead{$h_1$} & 
\colhead{$h_2$} & 
\colhead{$h_3$} &
\colhead{$k$}
}
\startdata
$\alpha$ & $-2.26 \times 10^{1}$ & $1.00$ & $-9.63 \times 10^{2}$ & $6.85 \times 10^{3}$ & $3.96 \times 10^{-3}$ & $-5.72 \times 10^{3}$ & $4.45 \times 10^{-2}$ \\
$\beta$ & $5.05 \times 10^{-1}$ & $9.09 \times 10^{-1}$ & $-1.84 \times 10^{2}$ & $1.84 \times 10^{2}$ & $3.93 \times 10^{-1}$ & $9.21 \times 10^{-1}$ & $3.92 \times 10^{-1}$ \\
$\log_{10}\dot{m}_\star^{\rm norm}$ & $5.02$ & $4.41 \times 10^{-1}$ & $1.02 \times 10^{1}$ & $-1.12 \times 10^{3}$ & $1.85 \times 10^{-3}$ & $1.10 \times 10^{3}$ & $1.14 \times 10^{-1}$ \\
$\log_{10}m_{\rm turn}$ & $3.23 \times 10^{1}$ & $1.03$ & $3.96 \times 10^{3}$ & $-4.19 \times 10^{3}$ & $1.27$ & $1.12 \times 10^{1}$ & $1.28$ \\
\enddata
\end{deluxetable*}

\section{Fitting the observed stellar mass function}\label{app:SMF_fit}
As described in Section~\ref{sec:obs_SMF}, we fit the \textit{observed} star-forming and quiescent mass functions with a double Schechter function (Eq.~(\ref{eq:dbl_schechter})) convolved with a lognormal stellar mass uncertainty (Eq.~(\ref{eq:SMF_uncertainty})), with the redshift evolution of this uncertainty characterized by the fit to a literature compilation described in \cite{rodriguez-puebla_matching_2025}. With this in hand, we run an MCMC with a Gaussian likelihood with the following priors on each parameter.

To enforce redshift continuity we parameterize each of these as cubic polynomial functions of redshift:
\begin{align}\label{eq:polynom_params}
\begin{split}
    p(z) &= p_{0} + p_{1}z + p_{2}z^2 + p_{3} z^3,
    \end{split}
\end{align}
where $p = \{\log_{10}\phi_1^*, \log_{10}\phi_2^*, \M^*, \alpha_1, \alpha_2\}$. 

We find that in order to achieve a satisfactory fit to the observed star-forming galaxy data, it is sufficient to use a single Schechter function with a cubic polynomial redshift evolution for each parameter. In this case, to limit some of the Schechter parameter degeneracies and motivated by \cite{peng_mass_2014a}, we enforce that the low-redshift ($z\leq 4$) faint-end slope is roughly fixed to $\alpha_1\sim -1.5$. Outside of that range, we adopt a uniform prior on $\alpha_1 \in [-3, -1]$, and over all redshifts, we allow $\M^*\in [9,12]$ and $\log_{10}\phi_1^*\in [-8,-2]$. Note that here we have adopted priors on the value of the parameters themselves ($p(z)$) at each redshift for which we compare our model to the data, rather than placing a prior on the polynomial coefficients themselves.

For the quiescent population --- which is only characterized between $0 \leq z \leq 5$ --- the data requires a double Schechter function, but because of this added flexibility (and comparatively limited redshift coverage), each parameter is fit with a linear or quadratic polynomial (as is found in \citealp{mcleod_evolution_2021}). We again adopt a uniform prior on the parameters, taking $p_0$ for each parameter to be $\log_{10}\phi_1^*\in [-5,-2]$, $\log_{10}\phi_2^*\in [-5,-2.5]$, $\M^*\in [10.4, 11]$, $\alpha_1\in [-2,0.5]$, and $\alpha_2\in [-3,0.5]$. We place a prior on the polynomial coefficients as well, enforcing that $p_1,p_2\in [-2, 2]$ for all parameters.

This procedure results in the converged posterior distributions reported in Table~\ref{tab:SMF_fit_params} and the fits shown in Figure~\ref{fig:SMF_fit}.

\begin{figure*}
    \centering
    \includegraphics[width=\linewidth]{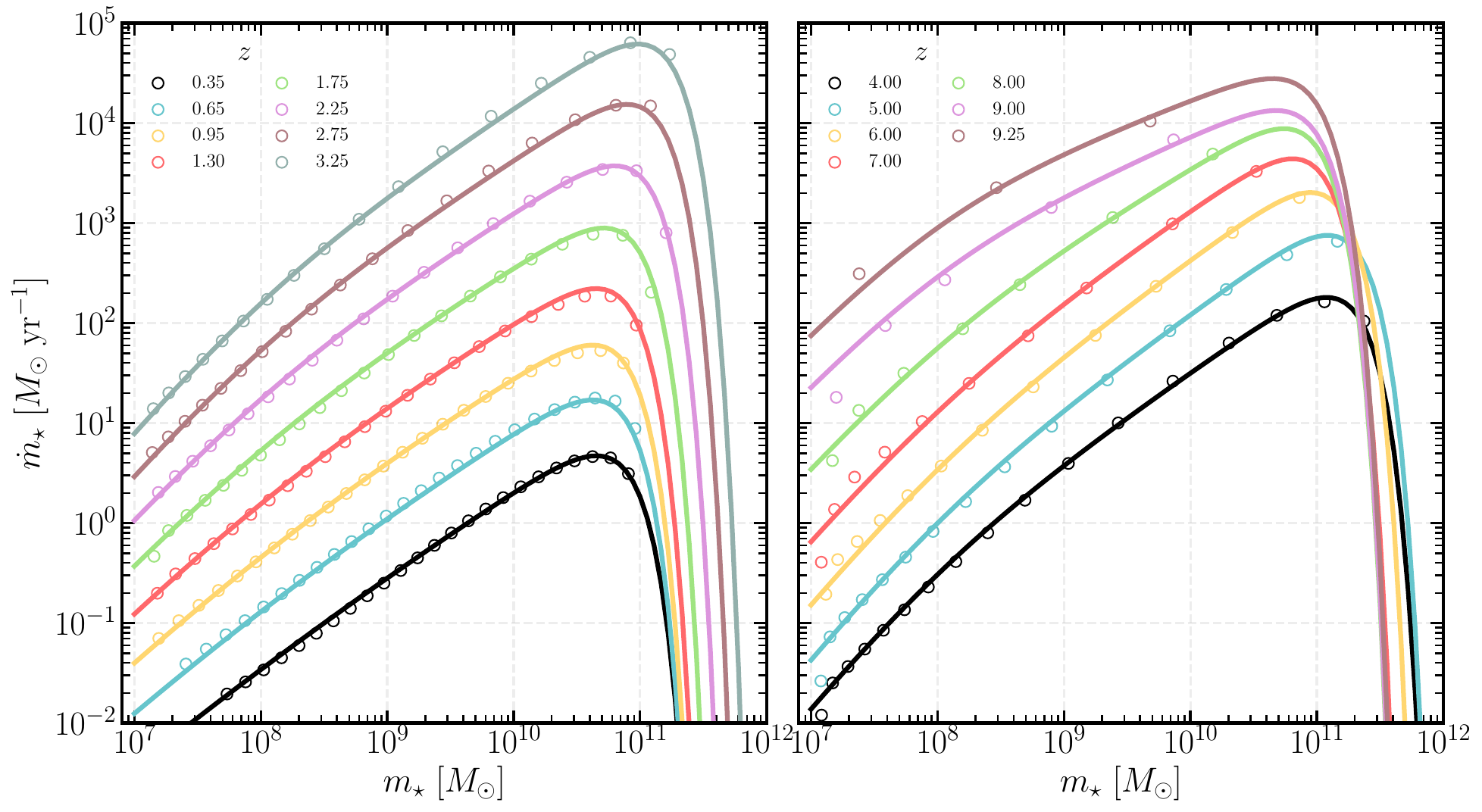}
    \caption{\textbf{The fitting function (Eqs.~(\ref{eq:MS_fit_dpl}), (\ref{eq:MS_fit_z_dep}), and (\ref{eq:logistic})) captures the redshift evolution and shape of the SMF-based star-forming MS very well.} The SFRs inferred from the continuity equation procedure outlined in Section~\ref{sec:continuity} (points) compared with the fitting formula presented in Section~\ref{sec:SMF_MS} (lines) at different redshifts (colors). The different redshifts have been offset by $+0.4$ dex and split into two panels for clarity.}
    \label{fig:MS_fit}
\end{figure*}

\section{Fitting the star-forming MS from the evolution of the SMF}\label{app:star-forming MS_fit}
As described in Section~\ref{sec:SMF_MS}, we fit the star-forming MS derived from the SMF with a double power law at low masses ($m_\star \lesssim 10^{10.5} M_\odot$) and an exponential cutoff at higher masses (Eq.~(\ref{eq:MS_fit_dpl})). This functional form is characterized by four parameters at each redshift: the high mass power law slope $\alpha$, the low mass power law slope $\beta$, the overall normalization $\dot{m}_\star^{\rm norm}$, and the turnover mass $m_{\rm turn}$. We use the \textsc{scipy} function \texttt{curve\_fit} to identify the best fit values of each of these parameters in each redshift bin from $z=0.2-9$.\footnote{We limit our fit to $z\gtrsim 0.2$ because discrepancies in the stellar mass inference procedure and survey areas between the surveys at the lowest redshifts lead to unphysical increases in the inferred galaxy SFRs at the latest times. As a result, the behavior of our fit below $z<0.2$ is an extrapolation, but we find that it nevertheless agrees well with the $z\sim 0$ results from other observed star-forming MS measurements.} 

From visual inspection of the evolution of the resulting parameters over this redshift range (points in Figure~\ref{fig:star-forming MS_params_zDep}), we find a piecewise redshift evolution --- a power law at $z\lesssim 6$ and an exponential at $z>6$, each of which is characterized by three parameters (Eq.~(\ref{eq:MS_fit_z_dep})):
\begin{equation}
    \label{eq:MS_fit_z_dep}
    \begin{split}
        r(z) = \sigma(&z|k,z_0) \big( h_1  e^{-h_2(z-z_{0})} + h_3 \big) \\
      &+ \big(1-\sigma(z|k,z_0)\big) \big( l_1 (1+z)^{l_2} + l_3 \big) \ ,
    \end{split}  
\end{equation}
where the $h_i$ ($l_i$) are the high- (low-)redshift parameters and $\sigma(z|k,z_0)$ is a logistic function that smoothly connects the two regimes. Here, the logistic function has a spread characterized by $k$ and a transition redshift of $z_0=6$:
\begin{equation}
    \label{eq:logistic}
    \sigma(z|k,z_0) = \frac{1}{1+e^{-k(z-z_0)}}\ .
\end{equation}
We fit this seven-parameter redshift evolution to each parameter again using the \textsc{scipy} module \texttt{curve\_fit}, resulting in the smooth curves shown in Figure~\ref{fig:star-forming MS_params_zDep} and the parameters listed in Table~\ref{tab:star-forming MS_fit_params}.

\clearpage

\bibliography{main}
\bibliographystyle{aasjournalv7}

\end{document}